# The emission probabilities of long range alpha particles from even-even $^{244-252}$Cm isotopes

K. P. Santhosh, Sreejith Krishnan and B. Priyanka

School of Pure and Applied Physics, Kannur University, Swami Anandatheertha Campus, Payyanur 670327, Kerala, INDIA

Email: drkpsanthosh@gmail.com

**Abstract.**

The alpha accompanied cold ternary fission of even-even $^{244}$Cm, $^{246}$Cm, $^{248}$Cm, $^{250}$Cm and $^{252}$Cm isotopes have been studied by taking the interacting barrier as the sum of Coulomb and proximity potential with the fragments in equatorial configuration. The favorable fragment combinations are obtained from the cold reaction valley plot and by calculating the relative yield for the charge minimized fragments. In the alpha accompanied ternary fission of $^{244}$Cm isotope, the highest yield is found for the fragment combination $^{110}$Ru+$^{4}$He+$^{130}$Sn, which possess near doubly magic nuclei $^{130}$Sn. For the ternary fission of $^{246}$Cm, $^{248}$Cm, $^{250}$Cm and $^{252}$Cm isotopes with $^{4}$He as light charged particle, the highest yield is obtained for the fragment combination with doubly magic nuclei $^{132}$Sn as the heavier fragment. The emission probabilities and kinetic energies of long range alpha particle have been computed for the $^{242,244,246,248}$Cm isotopes and are found to be in good agreement with the experimental data. The relative yields for the $^{4}$He accompanied ternary fission (equatorial and collinear) of $^{242-252}$Cm isotopes are compared with the corresponding yield for binary fission. The effect of deformation and orientation of fragments in the $^{4}$He accompanied ternary fission of $^{244-252}$Cm isotopes are studied. Our study reveals that the ground state deformation has an important role in the alpha accompanied ternary fission as that of shell effect.

**Keyword:** Spontaneous fission, Ternary fission, Heavy particle decay

**PACS Nos.** 25.85.Ca; 23.70.+j; 27.90.+b

## 1. Introduction

In 1947, Alvarez et al. [1] discovered the ternary fission with $^4$He as light charged particle in the fission of $^{235}$U, using photographic plate methods and using various counter techniques. Using the coincidence-counting methods, the emission of short-range particles in the fission was detected by Cassels et al. [2] and was found that, in the slow neutron induced fission of $^{235}$U, light charged particles were emitted in addition to the two heavy fragments which is approximately 4% of the fission events. Malkin et al. [3] have measured 400 cases of ternary fission of $^{244}$Cm with the emission of long range alpha particle, and have obtained the most probable energies for light and heavy fragments as (100 ± 8)MeV and (77 ± 6)MeV respectively.

Using a double folding potential, the isotopic yields have been studied for the alpha accompanied cold ternary fission of $^{248}$Cm by Sandulescu et al. [4-6]. The influence of the fragment excitation energies on the yields of $^{10}$Be accompanied cold ternary fission of $^{252}$Cf, including the level densities and the β-stretching of the fragments has also been studied. By taking the nuclear proximity force into account, the longitudinal ternary and binary fission barriers of $^{36}$Ar, $^{56}$Ni, and $^{252}$Cf has been studied by Royer et al. [7-9]. The authors found that for light nuclei, cascade fission is favored whereas the prolate fission is the most favorable for the heavier nuclei.

Within a stationary scattering formalism Delion et al. [10-12] provided a quantum description of the cold ternary fission process of $^4$He and $^{10}$Be accompanied cold ternary fission of $^{252}$Cf. The authors have also shown that the angular distribution of emitted light particle is strongly connected with its deformation and the equatorial distance. Gherghescu et al. [13] studied the decay of superheavy nuclei into two and three fragments using the macroscopic-microscopic method to obtain the deformation energy. The transition towards three equal fragment partitions was computed using the three centre shell model. Jandel et al. [14] studied the γ-ray multiplicity for the tripartition of $^{252}$Cf with He, Be and C as the third light charged particle. Andreev et al. [15] calculated the charge distributions for the ternary fragmentation of $^{252}$Cf and induced ternary fission of $^{56}$Ni. The relative yields for different light charged particle and also the mean total kinetic energy of fragments were also calculated. Mirzaei et al. [16] computed the deformation energy within the liquid drop model in the frame work of Yukawa plus exponential model and the penetrability for the binary process was compared with the ternary ones.

Florescu et al. [17] estimated the preformation amplitudes for $^4$He and $^{10}$Be clusters in the cold ternary fragmentation of $^{252}$Cf using a microscopic model starting from single particle spherical Woods-Saxon wave functions and with a large space BCS-type configuration mixing. A comparative study was done by Misicu et al. [18], for the three clusters with the triangular configuration, and with linear configuration where the alpha cluster is sandwiched between the heavier fragments. Ronen [19] calculated the absolute ternary particle emission probability for the isotopes $^{240-242}$Pu, $^{242-244}$Cm, $^{250-252}$Cf and $^{256-257}$Fm and suggested the ternary fission as a cluster decay of the fissioning nucleus in the last phase of the scission process.

Hamilton et al. [20] studied the cold ternary fragmentation of $^{252}$Cf using the triple gamma coincidence technique using Gammasphere with 72 gamma ray detectors in the case of $^4$He, $^{10}$Be and $^{14}$C as light charged particle. Ramayya et al. [21, 22] obtained the isotopic yields for the alpha ternary fission of $^{252}$Cf per 100 fission events and the highest yield is obtained for the fragment combination with $^{103}$Zr+$^{145}$Ba. The relative ternary yields of $^4$He, $^5$He and $^{10}$Be accompanied fission of $^{252}$Cf were also analyzed by the authors. The yields of long range alpha particle emitted during the spontaneous fission of $^{238, 240, 242, 244}$Pu was measured by Serot and Wagemans [23]. For tritons and α particles emitted in the spontaneous ternary fission of $^{244}$Cm, $^{246}$Cm, $^{248}$Cm, $^{250}$Cf and $^{252}$Cf and in the neutron induced fission of $^{243}$Cm, $^{245}$Cm $^{247}$Cm, $^{249}$Cf and $^{251}$Cf, the energy distributions and the emission probabilities were determined by Vermote et al. [24, 25].

As an extension of the preformed cluster model (PCM) of Gupta and collaborators, Manimaran and Balasubramaniam [26-29] developed a three cluster model (TCM) to study the ternary fragmentation of $^{252}$Cf for all possible fragmentation in the equatorial and collinear configuration. The deformation and orientation effects were also discussed with the $^4$He and $^{10}$Be accompanied ternary fission of $^{252}$Cf. Shagun et al. [30] studied the fragmentation potential for alpha accompanied ternary fission of superheavy nuclei with Z = 114, 116, 118 and 120.

Taking Coulomb and proximity potential [31] as the barrier the ternary fragmentation of $^{242}$Cm emitting light clusters $^4$He, $^{10}$Be and $^{14}$C have been studied with fragments in equatorial configuration. In the present work, we are interested in the study of alpha accompanied ternary fission of $^{244}$Cm, $^{246}$Cm, $^{248}$Cm, $^{250}$Cm and $^{252}$Cm isotopes and we have considered the equatorial configuration of fragments, as the experimental studies of Oertzen et al. [32] revealed that the equatorial configuration is the preferred configuration for the ternary fission accompanying light charged particles $^4$He, $^{10}$Be etc. Using TCM, Manimaran et al. [29] have found that the equatorial configuration to be the most preferred one than the collinear configuration for the ternary fission accompanying light charged particles. The schematic diagram for the touching configuration of three spherical fragments in equatorial configuration is shown in figure 1(a). We have also studied the relative yield in the alpha accompanied ternary fission of $^{242-252}$Cm isotopes with fragments in collinear configuration. The relative yields for the $^4$He accompanied ternary fission (equatorial and collinear) of $^{242-252}$Cm isotopes are compared with the corresponding yield for binary fission. The calculations are done by taking the interacting potential barrier as the sum of Coulomb and proximity potential. It is to be noted that the Coulomb and proximity potential have been extensively used by one of us (KPS) for the studies on alpha decay [33, 34], cluster decay [35] and heavy ion induced fusion [36]. The influence of deformation and orientation of fragments in the $^4$He accompanied ternary fission of $^{244-252}$Cm isotopes are also studied.

Here we would like to mention that $^{242-250}$Cm is an alpha emitter [37] and $^{34}$Si cluster has been found to be emitted from the $^{242}$Cm isotope [38]. Our study on heavy particle emission from various Cm isotopes [39] predicts $^{34}$Si cluster emission from $^{244-252}$Cm with half lives within the experimental limit for measurements. The isotopes $^{242,244,246,248}$Cm also exhibit alpha accompanied ternary fission [23, 24] and this is the reason for taking $^{242-252}$Cm isotopes for the present study.

The formalism used for our calculation is described in Section 2. The results and discussion on the $^4$He accompanied ternary fission of $^{244}$Cm, $^{246}$Cm, $^{248}$Cm, $^{250}$Cm and $^{252}$Cm isotopes in the equatorial and collinear configuration is given in Section 3 and we summarize the entire work in Section 4.

## 2. Unified Ternary Fission Model (UTFM)

The light charged particle accompanied ternary fission is energetically possible only if $Q$ value of the reaction is positive. ie.

$$Q = M - \sum_{i=1}^{3} m_i > 0 \qquad (1)$$

Here $M$ is the mass excess of the parent and $m_i$ is the mass excess of the fragments. The interacting potential barrier for a parent nucleus exhibiting cold ternary fission consists of

Coulomb potential and nuclear proximity potential of Blocki et al [40, 41]. The proximity potential was first used by Shi and Swiatecki [42] in an empirical manner and has been quite extensively used by Gupta et al., [43] in the preformed cluster model (PCM) and is based on pocket formula of Blocki et al [40]. But in the present manuscript, another formulation of proximity potential (eqn (21a) and eqn (21b) of Ref. [41]) is used as given by Eqs. 7 and 8. The interacting potential barrier is given by

$$V = \sum_{i}^{3} \sum_{j>i}^{3} (V_{Cij} + V_{Pij}) \qquad (2)$$

with $V_{Cij} = \dfrac{Z_i Z_j e^2}{r_{ij}}$, the Coulomb interaction between the fragments. Here $Z_i$ and $Z_j$ are the atomic numbers of the fragments and $r_{ij}$ is the distance between fragment centers. The nuclear proximity potential [40] between the fragments is

$$V_{Pij}(z) = 4\pi \gamma b \left[ \frac{C_i C_j}{(C_i + C_j)} \right] \Phi\left(\frac{z}{b}\right) \qquad (3)$$

Here $\Phi$ is the universal proximity potential and $z$ is the distance between the near surfaces of the fragments. The distance between the near surfaces of the fragments for equatorial configuration is considered as $z_{12} = z_{23} = z_{13} = z$ and for collinear configuration the distance of separation are $z_{12} = z_{23} = z$ and $z_{13} = 2(C_2 + z)$. In collinear configuration the second fragment is considered to lie in between the first and third fragment. The Süssmann central radii $C_i$ of the fragments related to sharp radii $R_i$ is,

$$C_i = R_i - \left(\frac{b^2}{R_i}\right) \qquad (4)$$

For $R_i$ we use semi empirical formula in terms of mass number $A_i$ as [40]

$$R_i = 1.28 A_i^{1/3} - 0.76 + 0.8 A_i^{-1/3} \qquad (5)$$

The nuclear surface tension coefficient called Lysekil mass formula [44] is

$$\gamma = 0.9517 [1 - 1.7826 (N - Z)^2 / A^2] \text{ MeV/fm}^2 \qquad (6)$$

where $N$, $Z$ and $A$ represents neutron, proton and mass number of the parent, $\Phi$, the universal proximity potential (eqn (21a) and eqn (21b) of Ref. [41]) is given as,

$$\Phi(\varepsilon) = -4.41e^{-\varepsilon/0.7176} \quad , \text{ for } \varepsilon > 1.9475 \tag{7}$$

$$\Phi(\varepsilon) = -1.7817 + 0.9270\varepsilon + 0.0169\varepsilon^2 - 0.05148\varepsilon^3 \text{ , for } 0 \leq \varepsilon \leq 1.9475 \tag{8}$$

with $\varepsilon = z/b$, where the width (diffuseness) of the nuclear surface $b \approx 1$ fermi.

Using one-dimensional WKB approximation, barrier penetrability $P$, the probability for which the ternary fragments to cross the three body potential barrier is given as

$$P = \exp\left\{-\frac{2}{\hbar}\int_{z_1}^{z_2}\sqrt{2\mu(V-Q)}dz\right\} \tag{9}$$

The turning points $z_1 = 0$ represent touching configuration and $z_2$ is determined from the equation $V(z_2) = Q$, where $Q$ is the decay energy. The potential V in eqn. 9, which is the sum of Coulomb and proximity potential given by eqn. 2, and are computed by varying the distance between the near surfaces of the fragments. In eqn. 9 the mass parameter is replaced by reduced mass $\mu$ and is defined as,

$$\mu = m\left(\frac{\mu_{12}A_3}{\mu_{12}+A_3}\right) \tag{10}$$

and
$$\mu_{12} = \frac{A_1 A_2}{A_1 + A_2} \tag{11}$$

where $m$ is the nucleon mass and $A_1$, $A_2$, $A_3$ are the mass numbers of the three fragments. The relative yield can be calculated as the ratio between the penetration probability of a given fragmentation over the sum of penetration probabilities of all possible fragmentation as follows,

$$Y(A_i, Z_i) = \frac{P(A_i, Z_i)}{\sum P(A_i, Z_i)} \tag{12}$$

Manimaran et al. [26] obtained a similar trend for the relative yield of various fragmentation channels in both the cases when the distance of separation between the fragments is equal and unequal [26]. This is the reason for taking the distance of separation between the fragments as equal.

### 3. Results and Discussions

The quantum mechanical fragmentation theory (QMFT) [45] is able to describe cold fusion, cold fission and cluster radioactivity from a unified point of view. The unifying aspect of this theory is the shell closure effects of one or both the reaction partners for fusion or that of the decay products for fission and cluster radioactivity. In QMFT, the role of shell effects (largest for a spherically closed or nearly closed shell nucleus) arises through "cold reaction" or "cold decay" valleys, corresponding to the potential energy minima in the calculated fragmentation potential.

The ternary fragmentation of $^{244}$Cm, $^{246}$Cm, $^{248}$Cm, $^{250}$Cm and $^{252}$Cm with $^4$He as light charged particle for the equatorial configuration is studied using the concept of cold reaction valley which was introduced in relation to the structure of minima in the so called driving potential. The driving potential is defined as the difference between the interaction potential V and the decay energy Q of the reaction. The relative yield is calculated by taking the interacting potential barrier as the sum of Coulomb and proximity potential. The Q values are calculated using the recent mass tables of Wang et al. [46]. The driving potential (V−Q) for the parent nucleus is calculated (keeping third fragment $A_3$ as fixed) for all possible fragments as a function of mass and charge asymmetries respectively given as $\eta = \dfrac{A_1 - A_2}{A_1 + A_2}$ and $\eta_Z = \dfrac{Z_1 - Z_2}{Z_1 + Z_2}$, at the touching configuration. For every fixed mass pair ($A_1$, $A_2$) a pair of charges is singled out for which driving potential is minimized.

The equatorial configuration of three fragments of equal mass, and two heavy and one light fragments are pictorially represented in Fig 1 of Ref [47]. It can be seen from the figure that the lines joining the fragment centers form a triangle (1) an equilateral triangle in former case and (2) an isosceles triangle in later case. But in the case of fragments with different masses (and in the case of 2 alpha particles and one heavy fragment) the lines joining the fragment centers form a triangle rather than an equilateral triangle or isosceles triangle and the third (light) fragment tends to move in a direction perpendicular to the line joining the centers of other two fragments, the perpendicular line is drawn at the point where the surfaces of the two fragment is in contact.

### 3.1. Alpha accompanied ternary fission of $^{244}$Cm

The driving potential is calculated for the cold ternary fission of $^{244}$Cm with $^4$He as light charged particle and is plotted as a function of mass number $A_1$ as shown in figure 2. In the cold valley, the minima is found for the fragment combinations with mass number $A_1=$ $^4$He, $^{10}$Be, $^{14}$C, $^{16}$C, $^{20}$O, $^{22}$O, $^{24}$O, $^{26}$Ne, $^{30}$Mg, $^{32}$Mg, $^{34}$Si, $^{36}$Si, $^{40}$S, $^{42}$S, $^{44}$Ar, $^{46}$Ar, $^{48}$Ar, $^{50}$Ca, $^{52}$Ca etc. The deepest minimum is found for the fragment combination $^4$He+$^4$He+$^{236}$U. The other minima valleys are found around the fragment combinations $^{82}$Ge+$^{4}$He+$^{158}$Sm and $^{106}$Mo+$^{4}$He+$^{134}$Te. Of these, the fragment combinations $^{106}$Mo+$^{4}$He+$^{134}$Te with higher $Q$ values will be the most favorable fragments for the alpha accompanied ternary fission of $^{244}$Cm and is due to the presence of near doubly magic nuclei $^{134}$Te (N=82, Z=52)

The barrier penetrability is calculated for each charge minimized fragments in the cold ternary fission of $^{244}$Cm using the formalism described above. The relative yield is calculated and plotted it as a function of mass numbers $A_1$ and $A_2$ as shown in figure 3(b). The highest yield is obtained for the fragment combination $^{110}$Ru+$^{4}$He+$^{130}$Sn, which is due to the presence of near doubly magic nuclei $^{130}$Sn (N=80, Z=50) and the highest $Q$ value of 216.233MeV. The next highest yield found for the fragment combinations $^{106}$Mo+$^{4}$He+$^{134}$Te is due to the near doubly magic nuclei $^{134}$Te (N=82, Z=52) and the yield obtained for the fragment combination $^{108}$Ru+$^{4}$He+$^{132}$Sn is due to the presence of doubly magic nuclei $^{132}$Sn (N=82, Z=50). The closed shell effect Z=50 of $^{128}$Sn and $^{126}$Sn makes the fragment combinations $^{112}$Ru+$^{4}$He+$^{128}$Sn and $^{114}$Ru+$^{4}$He+$^{126}$Sn with relative higher yield. For a better comparison of the yield, a histogram is plotted as a function of mass numbers $A_1$ and $A_2$ as shown in figure 4. The hatched bars belong to odd mass numbers and the dark ones belong to even mass numbers.

## 3.2. Alpha accompanied ternary fission of $^{246}$Cm

For the alpha accompanied ternary fission of $^{246}$Cm, the driving potential is calculated and plotted as a function of mass number $A_1$ as shown in figure 11, denoted as spherical. The minima is found for the fragment combinations with mass number $A_1=$ $^{4}$He, $^{10}$Be, $^{14}$C, $^{16}$C, $^{18}$C, $^{20}$O, $^{22}$O etc. The deepest minimum is found for the fragment combination $^{4}$He+$^{4}$He+$^{238}$U. Moving on the fission region three deep valleys are found; one around $^{36}$Si+$^{4}$He+$^{206}$Hg which possess near doubly magic nuclei $^{206}$Hg, second one around $^{82}$Ge+$^{4}$He+$^{160}$Sm possessing the neutron closed shell N=50 of $^{82}$Ge and the third one around $^{110}$Ru+$^{4}$He+$^{132}$Sn which possess near doubly magic nuclei $^{132}$Sn (N=82, Z=50) and also higher $Q$ value.

The relative yield is calculated and plotted it as a function of mass numbers $A_1$ and $A_2$ as shown in figure 3(c). The highest yield is obtained for $^{110}$Ru+$^{4}$He+$^{132}$Sn which possess doubly magic nuclei $^{132}$Sn (N=82, Z=50) and highest $Q$ value of 216.808MeV. The presence of near doubly magic nuclei $^{130}$Sn (N=80, Z=50) and $^{134}$Te (N=82, Z=52) respectively makes the fragment combinations $^{112}$Ru+$^{4}$He+$^{130}$Sn and $^{108}$Mo+$^{4}$He+$^{134}$Te with relatively higher yield. The high yield found for the splitting $^{114}$Ru+$^{4}$He+$^{128}$Sn is due to the proton magic number Z=50 of

$^{128}$Sn and the presence of near doubly magic nuclei $^{136}$Te (N=84, Z=52) is to be quoted as the reason for the high yield obtained for $^{106}$Mo+$^{4}$He+$^{136}$Te.

### 3.3. Alpha accompanied ternary fission of $^{248}$Cm

The driving potential is calculated for the alpha accompanied ternary fission of $^{248}$Cm and plotted it as a function of mass number $A_1$ as shown in figure 12, denoted as spherical. In the cold valley the minima is found for the fragment combination with $A_1=$ $^{4}$He, $^{10}$Be, $^{14}$C, $^{16}$C, $^{18}$C, $^{20}$O, $^{22}$O etc. The deepest minimum is found for the fragment combination $^{4}$He+$^{4}$He+$^{240}$U, possesses $Q$ value of 9.8271MeV. The other deep valleys are found around the fragment combinations $^{38}$Si+$^{4}$He+$^{206}$Hg which possess near doubly magic nuclei $^{206}$Hg (N=126, Z=80) and $^{82}$Ge+$^{4}$He+$^{162}$Sm which possess closed shell effect Z=50 of $^{82}$Ge. The fragment combination occur around $^{112}$Ru+$^{4}$He+$^{132}$Sn with higher $Q$ value of 217.141MeV may be the most favorable fragments occur in the $^{4}$He accompanied ternary fission of $^{248}$Cm which is due to the presence of doubly magic nuclei $^{132}$Sn.

The relative yield is calculated and plotted it as a function of mass number $A_1$ as shown in figure 3(d). The highest yield is obtained for the fragment combination $^{112}$Ru+$^{4}$He+$^{132}$Sn which possess doubly magic nuclei $^{132}$Sn (N=82, Z=50) and highest $Q$ value of 217.141MeV. The next highest yield obtained for the splitting $^{114}$Ru+$^{4}$He+$^{130}$Sn and $^{110}$Mo+$^{4}$He+$^{134}$Te is due to the presence of near doubly magic nuclei $^{130}$Sn (N=80, Z=50) and $^{134}$Te (N=82, Z=52) respectively and the yield obtained for $^{116}$Ru+$^{4}$He+$^{128}$Sn is due to the closed shell effect of $^{128}$Sn (Z=50). The presence of near doubly magic nuclei $^{136}$Te makes the fragment combination $^{108}$Mo+$^{4}$He+$^{136}$Te with relative higher yield.

### 3.4. Alpha accompanied ternary fission of $^{250}$Cm

The driving potential is calculated for the alpha accompanied ternary fission of $^{250}$Cm and plotted it as a function of mass number $A_1$ as shown in figure 13, denoted as spherical. The minima obtained in the cold valley for the fragment combinations with mass number $A_1=$ $^{4}$He, $^{10}$Be, $^{14}$C, $^{16}$C, $^{18}$C, $^{20}$O, $^{22}$O etc. The deepest minimum valley is obtained for the fragment combination $^{4}$He+$^{4}$He+$^{242}$U which possess a low $Q$ value of 9.520MeV. Moving on the fission region three deep valleys are found one around for the fragment combination $^{42}$S+$^{4}$He+$^{204}$Pt which possess neutron closed shell of $^{204}$Pt (N=126), second one around the fragment combination $^{80}$Zn+$^{4}$He+$^{166}$Gd possessing the neutron closed shell of $^{80}$Zn (N=50) and the third one around $^{114}$Ru+$^{4}$He+$^{132}$Sn. The fragment combination $^{114}$Ru+$^{4}$He+$^{132}$Sn possess the presence of doubly magic nuclei $^{132}$Sn (N=82, Z=50).

The relative yield is calculated and plotted it as a function of mass numbers $A_1$ and $A_2$ as shown in figure 3(e). The highest yield obtained for the fragment combinations $^{110}$Mo+$^{4}$He+$^{136}$Te and $^{112}$Mo+$^{4}$He+$^{134}$Te are due to the nearly doubly magic nuclei $^{136}$Te (N=84, Z=52) and $^{134}$Te (N=82, Z=52) respectively. The highest yield is obtained for the fragment combination $^{114}$Ru+$^{4}$He+$^{132}$Sn which possess doubly magic nuclei $^{132}$Sn (N=82, Z=50) and highest $Q$ value 217.331MeV. The next highest yield obtained for the fragment combination $^{116}$Ru+$^{4}$He+$^{130}$Sn is due to the presence of near doubly magic nuclei $^{130}$Sn (N=80, Z=50) and the yield obtained for the fragment combination $^{118}$Pd+$^{4}$He+$^{128}$Cd is due to the presence of neutron closed shell N=80 of $^{128}$Cd.

### 3.5. Alpha accompanied ternary fission of $^{252}$Cm

The driving potential is calculated for the alpha accompanied ternary fission of $^{252}$Cm and plotted it as a function of mass number $A_1$ as shown in figure 14, denoted as spherical. In the cold valley the minima is found for the fragment combination with $A_1$= $^{4}$He, $^{6}$He, $^{10}$Be, $^{12}$Be, $^{14}$C, $^{16}$C, $^{18}$C, $^{20}$O, $^{22}$O etc. The deepest minimum is found for the fragment combination with $^{4}$He+$^{4}$He+$^{244}$U. The other minimum valleys are found around the fragment combination $^{50}$Ca+$^{4}$He+$^{198}$W (possess proton shell closure Z=20 of $^{50}$Ca and near neutron shell closure N=124 of $^{198}$W) and $^{80}$Zn+$^{4}$He+$^{168}$Gd (possess neutron shell closure N=50 of $^{80}$Zn). The deep minimum valley occur at the fragment combination $^{116}$Ru+$^{4}$He+$^{132}$Sn possess doubly magic nuclei $^{132}$Sn and highest $Q$ value of 217.248MeV may be the probable fragment combination with highest yield.

The relative yield is calculated for each charge minimized third fragment and plotted it as a function of mass number $A_1$ and $A_2$ as shown in figure 3(f). The highest yield is found for the fragment combination $^{116}$Pd+$^{4}$He+$^{132}$Sn is due to the doubly magic nuclei $^{132}$Sn and highest $Q$ value of 217.248MeV. The next highest yield is found for the fragment combination $^{118}$Ru+$^{4}$He+$^{130}$Sn which possess near doubly magic nuclei $^{130}$Sn (N=80, Z=50) and the yield obtained for the fragment combinations $^{112}$Mo+$^{4}$He+$^{136}$Te and $^{114}$Ru+$^{4}$He+$^{134}$Sn are due to the near doubly magic nuclei $^{136}$Te (N=84, Z=52) and $^{134}$Sn (N=84, Z=50) respectively.

In figure 3, the relative yield is plotted for all the considered isotopes of curium as a function of mass numbers $A_1$ and $A_2$. The most probable fragment combinations occur with the alpha accompanied ternary fission is labeled. The relative yield obtained for the isotope $^{242}$Cm is also included in the figure. The mass numbers are labeled in the X-axis with $A_1>80$ and $A_2<168$ as the fragment combination possess higher Q values between the corresponding range of mass numbers. In all the cases, the plot of relative yield with mass number shows a two humped structure. From the figure, it is clear that the relative yield obtained in each isotope of curium increases with the increasing mass number.

### 3.6. Emission Probability of long range alpha particle

The emission probability of long range alpha particle (LRA) is determined with the number of fission events B, and the usual notation for the emission probability is LRA/B. Carjan [48] suggests that LRA emission is possible only if the α cluster is formed inside the fissioning nucleus and should gain enough energy to overcome the Coulomb barrier of the scission nucleus. Serot and Wagemans [23] demonstrated that the emission probability of long range alpha particle is strongly dependent on the spectroscopic factor or α cluster preformation factor $S_\alpha$, which can be calculated in a semi-empirical way proposed by Blendowske et al. [49] as, $S_\alpha = b\lambda_e / \lambda_{WKB}$ where $b$ is the branching ratio for the ground state to ground state transition, $\lambda_e$ is the experimental α decay constant and $\lambda_{WKB}$ is the α decay constant calculated from the WKB approximation. Vermote et al. [24] proved that $^4$He emission probability in spontaneous fission is about 20% higher than for neutron induced fission. The absolute emission probability is given by;

$$\frac{LRA}{B} = S_\alpha P_{LRA} \qquad (13)$$

With $P_{LRA}$ as the probability of the alpha particle when it is already present in fissioning nucleus given as,

$$P_{LRA} = \exp\left\{-\frac{2}{\hbar}\int_{z_0}^{z_1}\sqrt{2\mu(V-Q)}dz\right\} \qquad (14)$$

Here the first turning point is determined from the equation $V(z_0)=Q$, where $Q$ is the decay energy, and the second turning point $z_1=0$ represent the touching configuration. For the internal (overlap) region, the potential is taken as a simple power law interpolation. Here we have computed the emission probabilities of long range alpha particle in the case of $^{242}$Cm, $^{244}$Cm, $^{246}$Cm and $^{248}$Cm. The obtained results are found to be in good agreement with the experimental data [23, 24]. The spectroscopic factors and corresponding emission probabilities of $^{242-252}$Cm isotopes are listed in table 1.

### 3.7 Kinetic energies of long range $\alpha$ particle in the ternary fission of $^{242-252}$Cm isotopes

The kinetic energy of long range $\alpha$ particle emitted in the ternary fission of $^{242-252}$Cm isotopes are computed using the formalism reported by Fraenkel [50]. The conservation of total momentum in the direction of light particle and in a direction perpendicular to light particle leads to the relations,

$$(m_L E_L)^{1/2} = (m_H E_H)^{1/2} \cos\theta_R - (m_\alpha E_\alpha)^{1/2} \cos\theta_L \qquad (15)$$

$$(m_H E_H)^{1/2} \sin\theta_R = (m_\alpha E_\alpha)^{1/2} \sin\theta_L \qquad (16)$$

Here $m_L$, $m_H$ and $m_\alpha$ are the masses of the light, heavy and the $\alpha$ particle respectively. $E_L$, $E_H$ and $E_\alpha$ represent the final energies of the light, heavy and the $\alpha$ particle respectively. The kinetic energy of the long range alpha particle can be derived from eqn. (15) and eqn. (16) and is given as,

$$E_\alpha = E_L \left(\frac{m_L}{m_\alpha}\right) (\sin\theta_L \cot\theta_R - \cos\theta_L)^{-2} \qquad (17)$$

Here $\theta_L$ is the angle between the alpha particle and the light particle and $\theta_R$ is the recoil angle. The kinetic energy of light fragment $E_L$ is related to the total kinetic energies of fission fragments $TKE$ as

$$E_L = \frac{A_H}{A_L + A_H} TKE \qquad (18)$$

The total kinetic energies of fission fragments $TKE$ can be computed using the expressions reported by Viola et al. [51] or can be taken from Herbach et al. [52]. In the present work we have used the expression taken from Herbach et al. [52] given as

$$TKE = \frac{0.2904 (Z_L + Z_H)^2}{A_L^{1/3} + A_H^{1/3} - (A_L + A_H)^{1/3}} \frac{A_L A_H}{(A_L + A_H)^2} \qquad (19)$$

Here $A_L$ and $A_H$ are the mass numbers of light and heavy fragments respectively. Using the formalism described above we have computed the energy of long range alpha particle emitted from $^{242-252}$Cm isotopes for various fragmentation channels and is tabulated in table 2, where we have given the fragmentation channel and kinetic energy of the long range alpha particle $E_\alpha$. The computed $TKE$ values are found to be around 170MeV and according to Fraenkel [50], for the mean total energies of fission fragments $(\approx 168\,MeV)$, the maximum value of the recoil angle $\theta_R = 4.5^0$, and this maximum value is obtained for $\theta_L = 92.25^0$. For this reason, in the present manuscript we have taken $\theta_R = 4.5^0$ and $\theta_L = 92.25^0$. The experimental kinetic energy [24, 53] of the long range alpha particle in the ternary fission of $^{242,244,246,248}$Cm is also given in the table and has been compared with our calculated values. It is to be noted that, our predicted values are in good agreement with the experimental kinetic energies.

### 3.8. Alpha accompanied ternary fission of $^{242-252}$Cm in collinear configuration

The ternary fission of $^{242-252}$Cm isotopes with fragments in the collinear configuration are studied using the concept of cold reaction valley. In the collinear configuration, the light charged particle $^4$He ($A_2$) is considered in between the other two fragments. The schematic diagram for the touching configuration of three spherical fragments in collinear configuration is shown in figure 1(b). The driving potential $(V-Q)$ for each parent nuclei $^{242-252}$Cm is calculated and plotted as a function of mass number $A_1$ and is shown in figure 5. The fragment combinations with minimum driving potential (with high Q value) usually possess a higher relative yield. In the ternary fission of $^{242,\,244}$Cm isotopes, the fragment combinations $^{104}$Mo+$^4$He+$^{134}$Te and $^{110}$Ru+$^4$He+$^{130}$Sn which possess near doubly magic nuclei $^{134}$Te (N=82, Z=52) and $^{132}$Sn (N=80, Z=50) respectively, may have higher yields which can be verified through the calculation of penetrability. For the parent nuclei $^{246,\,248,\,250,\,252}$Cm, the minimum occurs for the fragment combination with the isotopes of $^{110,\,112,\,114,\,116}$Ru and doubly magic $^{132}$Sn (N=82, Z=50) respectively.

The barrier penetrability is calculated for all fragment combinations in the ternary fission of the parent nuclei $^{242}$Cm using eqn.(9) and hence the yield is calculated and plotted as a function of mass numbers $A_1$ and $A_3$ as shown in figure 6(a). The highest yield is obtained for the fragment combination $^{104}$Mo+$^4$He+$^{134}$Te, which possesses near doubly magic nuclei $^{134}$Te (N=82, Z=52). The next higher relative yield is obtained for the fragment combinations $^{106}$Mo+$^4$He+$^{132}$Te, $^{108}$Ru+$^4$He+$^{130}$Sn and $^{110}$Ru+$^4$He+$^{128}$Sn, due to the presence of near doubly magic nuclei $^{132}$Te (N=80, Z=52), near doubly magic nuclei $^{130}$Sn (N=80, Z=50) and the proton magicity of $^{128}$Sn (Z = 50) respectively.

For the ternary fission of $^{244}$Cm, the relative yield is plotted as a function of mass numbers $A_1$ and $A_3$ as shown in figure 6(b). The highest yield is obtained for the fragment combination $^{110}$Ru+$^{4}$He+$^{130}$Sn which possesses near doubly magic nuclei $^{130}$Sn (N=80, Z=50). The next higher relative yield found for the fragment combinations $^{106}$Mo+$^{4}$He+$^{134}$Te, $^{108}$Ru+$^{4}$He+$^{132}$Sn and $^{112}$Ru+$^{4}$He+$^{128}$Sn is due to the presence of near double magicity of $^{134}$Te (N=82, Z=52), doubly magic nuclei $^{132}$Sn (N=82, Z=50) and the proton magicity of $^{128}$Sn (Z = 50) respectively.

In figure 6(c), the relative yield obtained in the ternary fission of $^{246}$Cm is plotted as a function of mass numbers $A_1$ and $A_3$, in which the highest yield found for the fragment combination $^{110}$Ru+$^{4}$He+$^{132}$Sn which is due to the presence of doubly magic nuclei $^{132}$Sn (N=82, Z=50). The next higher yield obtained for the fragment combination $^{108}$Mo+$^{4}$He+$^{134}$Te, $^{112}$Ru+$^{4}$He+$^{130}$Sn and $^{114}$Ru+$^{4}$He+$^{128}$Sn is due to the presence of near doubly magic nuclei $^{134}$Te (N=82, Z=52), near double magicity of $^{130}$Sn (N=80, Z=50) and the presence of proton magic number Z=50 of $^{128}$Sn respectively.

Figures 6(d) - 6(f) represent the relative yield versus mass numbers $A_1$ and $A_3$ for the ternary fission of $^{248}$Cm - $^{252}$Cm respectively. In figure 6(d), the highest yield obtained for the fragment combination $^{112}$Ru+$^{4}$He+$^{132}$Sn is due to the doubly magic nuclei $^{132}$Sn (N=82, Z=50). The relative yield obtained for the fragment combinations $^{110}$Mo+$^{4}$He+$^{134}$Te and $^{114}$Ru+$^{4}$He+$^{130}$Sn are due to the near double magicity of $^{134}$Te (N=82, Z=52) and $^{130}$Sn (N=80, Z=50) respectively. From the figure 6(e) it is clear that, the highest yield is obtained for the fragment combination $^{114}$Ru+$^{4}$He+$^{132}$Sn which is due to the presence of doubly magic nuclei $^{132}$Sn (N=82, Z=50). The yield found for the fragment combination $^{116}$Ru+$^{4}$He+$^{130}$Sn is due to the presence of near doubly magic nuclei $^{130}$Sn (N=80, Z=50). In the case of $^{252}$Cm as shown in figure 6(f), the highest yield is obtained for the fragment combination $^{116}$Ru+$^{4}$He+$^{132}$Sn which possesses doubly magic nuclei $^{132}$Sn (N=82, Z=50). The next higher yield found for the fragment combination $^{118}$Ru+$^{4}$He+$^{130}$Sn is due to the near double magicity of $^{130}$Sn (N=80, Z=50).

Our study on ternary fragmentation of $^{242-252}$Cm with fragments in collinear configuration reveals the role of doubly magic nuclei $^{132}$Sn (N=82, Z=50), near double magicity of $^{130}$Sn (N=80, Z=50) and $^{134}$Te (N=82, Z=52). From the comparison of figure 3 and figure 6 for respective parent nuclei, it can be seen that the fragment combinations with highest yield obtained in the equatorial and collinear configurations are found to be the same. Also it should be noted that yield for equatorial configuration is twice as that of the collinear configuration and this reveals that equatorial configuration is the preferred configuration than collinear configuration in light charged particle accompanied ternary fission.

### 3.9. Binary fission of $^{242-252}$Cm isotopes

The driving potential for the binary fission of $^{242-252}$Cm is calculated using the concept of cold reaction valley and plotted as a function of mass number $A_1$ as shown in figure 7. In the binary fission of $^{242}$Cm and $^{244}$Cm, the fragment combinations occurring around $^{108}$Ru+$^{4}$He+$^{134}$Te and $^{110}$Ru+$^{4}$He+$^{134}$Te respectively have the minimum value of (V-Q) due to the presence of the near doubly magic $^{134}$Te (N=82, Z=52). For the binary fission of $^{246-252}$Cm isotopes, the fragment combinations found around the doubly magic nuclei $^{132}$Sn (N=82, Z=50) have higher relative yield which can be verified through the calculation of penetrability.

The barrier penetrability is calculated for all the possible binary fragmentations of $^{242}$Cm and hence the yield is calculated and plotted as a function of mass numbers $A_1$ and $A_2$ as shown in figure 8(a). The highest yield is obtained for the fragment combination $^{108}$Ru+$^{134}$Te which possesses the near doubly magic nuclei $^{134}$Te (N=82, Z=52). The yield found for the splitting $^{110}$Ru+$^{132}$Te, $^{112}$Pd+$^{130}$Sn and $^{114}$Pd+$^{128}$Sn is due to the presence of near double magicity of $^{132}$Te (N=80, Z=52), near doubly magic nuclei $^{130}$Sn (N=82, Z=50) and the closed shell effect Z=50 of $^{128}$Sn respectively.

In the $^{244}$Cm binary fragmentation, the yield is calculated and plotted as a function of mass numbers $A_1$ and $A_2$ as shown in figure 8(b). The highest yield is obtained for the fragment combination $^{110}$Ru+$^{134}$Te which possesses near doubly magic nuclei $^{134}$Te (N=82, Z=52). The yield found for the fragment combination $^{112}$Pd+$^{132}$Sn, $^{114}$Pd+$^{130}$Sn and $^{116}$Pd+$^{128}$Sn is due to the presence of doubly magic nuclei $^{132}$Sn (N=82, Z=50), near doubly magic nuclei $^{130}$Sn (N=80, Z=50) and the proton magicity of $^{128}$Sn (Z = 50) respectively.

For the binary fission of $^{246}$Cm as shown in figure 8(c), the highest relative yield is found for the fragment combination $^{114}$Pd+$^{132}$Sn which possesses the presence of doubly magic nuclei $^{132}$Sn (N=82, Z=50). The yield found for the fragment combination $^{112}$Ru+$^{134}$Te, $^{116}$Pd+$^{130}$Sn and $^{118}$Pd+$^{128}$Sn is due to the presence of near double magicity of $^{134}$Te (N=82, Z=52), $^{130}$Sn (N=80, Z=50) and proton shell effect Z=50 of $^{128}$Sn respectively.

Figures 8(d) - 8(f) represent the relative yield versus mass numbers $A_1$ and $A_2$ for the binary fission of $^{248}$Cm - $^{252}$Cm respectively. In figure 8(d), the highest yield is found for the fragment combination $^{116}$Pd+$^{132}$Sn which possesses doubly magic nuclei $^{132}$Sn (N=82, Z=50).

The next higher yield is found for the fragment combination $^{118}$Pd+$^{130}$Sn, which also possesses near double magicity of $^{130}$Sn (N=80, Z=50). In the binary fission of $^{250}$Cm as shown in figure8(e) the highest yield is found for the fragment combination $^{118}$Pd+$^{132}$Sn, which is due to the presence of doubly magic nuclei $^{132}$Sn. The next higher yield is found for the fragment combination $^{120}$Pd+$^{130}$Sn, which possesses near double magicity of $^{130}$Sn (N=80, Z=50). In the case of $^{252}$Cm, the highest relative yield is obtained for the fragment combination $^{120}$Pd+$^{132}$Sn which possesses doubly magic nuclei $^{132}$Sn (N=82, Z=50).

The relative yields obtained for the binary fission of $^{242-252}$Cm isotopes are compared with that of ternary fission (both the equatorial and collinear configuration) and plotted in figure9 as a bar graph. It can be seen that the yield obtained for the equatorial configuration is higher than that of the collinear configuration. From the figure it can also be seen that the yield for binary fission is higher than that of ternary fission (both equatorial and collinear configuration). This indicates to the fact that the probability for the occurrence of binary fragmentation is higher than that of ternary fragmentation and ternary fragmentation is observed 1 in 500 binary fissions.

### 3.10 Effect of deformation

The effect of deformation and orientation of fragments in $^4$He accompanied ternary fission of $^{244-252}$Cm isotopes have been analyzed taking the Coulomb and proximity potential as the interacting barrier. The Coulomb interaction between the two deformed and oriented nuclei, which is taken from [54] and which includes higher multipole deformation [55, 56], is given as,

$$V_C = \frac{Z_1 Z_2 e^2}{r} + 3Z_1 Z_2 e^2 \sum_{\lambda, i=1,2} \frac{1}{2\lambda+1} \frac{R_{0i}^\lambda}{r^{\lambda+1}} Y_\lambda^{(0)}(\alpha_i) \left[ \beta_{\lambda i} + \frac{4}{7} \beta_{\lambda i}^2 Y_\lambda^{(0)}(\alpha_i) \delta_{\lambda,2} \right] \quad (20)$$

with

$$R_i(\alpha_i) = R_{0i} \left[ 1 + \sum_\lambda \beta_{\lambda i} Y_\lambda^{(0)}(\alpha_i) \right] \quad (21)$$

where $R_{0i} = 1.28 A_i^{1/3} - 0.76 + 0.8 A_i^{-1/3}$. Here $\alpha_i$ is the angle between the radius vector and symmetry axis of the $i^{th}$ nuclei (see Fig.1 of Ref [55]) and it is to be noted that the quadrupole interaction term proportional to $\beta_{21}\beta_{22}$, is neglected because of its short range character.

In proximity potential, $V_P(z) = 4\pi\gamma b \bar{R} \Phi(\varepsilon)$, the deformation comes only in the mean curvature radius. For spherical nuclei, mean curvature radius is defined as $\bar{R} = \frac{C_1 C_2}{C_1 + C_2}$, where

$C_1$ and $C_2$ are Süssmann central radii of fragments. The mean curvature radius, $\overline{R}$ for two deformed nuclei lying in the same plane can be obtained by the relation,

$$\frac{1}{\overline{R}^2} = \frac{1}{R_{11}R_{12}} + \frac{1}{R_{21}R_{22}} + \frac{1}{R_{11}R_{22}} + \frac{1}{R_{21}R_{12}} \tag{22}$$

The four principal radii of curvature $R_{11}$, $R_{12}$, $R_{21}$ and $R_{22}$ are given by Baltz and Bayman [57].

Figures 10-14 represent the cold valley plots, the plot connecting the driving potential $(V-Q)$ and the mass number $A_1$ for $^{244}$Cm to $^{252}$Cm isotopes. In these plots three cases are considered (1) three fragments taken as spherical (2) two fragments ($A_1$ and $A_2$) as deformed with $0^0-0^0$ orientation and (3) two fragments ($A_1$ and $A_2$) as deformed with $90^0-90^0$ orientation. For computing driving potential we have used experimental quadrupole deformation ($\beta_2$) values taken from Ref. [58] and for the cases for which there are no experimental values, we have taken them from Moller et al [59]. It can be seen from these plots that in most of the cases, $0^0-0^0$ orientation have a low value for driving potential, but in few cases, $90^0-90^0$ orientation has the low value. In the former case, either both the fragments are prolate or one fragment is prolate and the other one is spherical; and in latter case both fragments are either oblate or one fragment is oblate and the other one is spherical. It can be seen that when deformation are included, the optimum fragment combination are also found to be changed. For e.g in the case of $^{244}$Cm the fragment combination $^{112}$Ru+$^{4}$He+$^{128}$Sn are changed to $^{112}$Pd+$^{4}$He+$^{128}$Cd when deformation is included; and in the case of $^{248}$Cm the fragment combination $^{116}$Ru+$^{4}$He +$^{128}$Sn changed to $^{116}$Pd+$^{4}$He +$^{128}$Cd with the inclusion of deformation.

We have computed barrier penetrability for all fragment combinations in the cold valley plot (figures 10-14) which have the minimum $(V-Q)$ value, with including the quadrupole deformation. The computations are done using the deformed Coulomb potential (eqn.20) and deformed nuclear proximity potential (eqn.22). Inclusion of quadrupole deformation ($\beta_2$) reduces the height and width of the barrier and as a result, the barrier penetrability is found to increase. For e.g. in the case of $^{244}$Cm, the fragment combination $^{86}$Br+$^{4}$He+$^{154}$Pr have barrier penetrability $P^{spherical}=4.04\times10^{-14}$ when treated as spherical and $P^{deformed}=4.96\times10^{-12}$ when deformation of fragments are included; and in the case of $^{246}$Cm the fragment combination $^{110}$Tc+$^{4}$He+$^{132}$Sb have barrier penetrability $P^{spherical}=2.28\times10^{-11}$ when treated as spherical and

$P^{deformed} = 3.09 \times 10^{-10}$ when deformation of fragments are included. It is to be noted that both the fragments are prolate deformed {$^{86}$Br ($\beta_2 = 0.071$), $^{154}$Pr ($\beta_2 = 0.27$)} in the former case and oblate deformed {$^{110}$Tc ($\beta_2 = -0.258$), $^{132}$Sb ($\beta_2 = -0.026$)} in the latter case. The relative yield is calculated and Figures 15-16 represent the plot connecting relative yield versus fragment mass number $A_1$ and $A_2$ for $^{244}$Cm to $^{252}$Cm. By comparing the figures 15-16 with corresponding plots for the spherical case, equatorial configuration (Fig 3) and collinear configuration (Fig 6), it can be seen that fragments with highest yield are also found to be changed. For e.g. in the case of $^{244}$Cm the fragments with highest yield are $^{110}$Ru and $^{130}$Sn when fragments are treated as spheres, but when deformation are included, the highest yield is found for the fragments $^{116}$Pd and $^{124}$Cd. For a better comparison of the result, a histogram is plotted with yield as a function of mass numbers $A_1$ and $A_2$ for the ternary fragmentation of $^{244-252}$Cm isotopes with the inclusion of quadrupole deformation $\beta_2$ as shown in figure 17-21. The studies on the influence of deformation in the alpha accompanied ternary fission of $^{244-252}$Cm isotopes reveal that the ground state deformation has an important role in ternary fission as that of shell effect.

Vermote et al. [24] experimentally studied the emission probability and the energy distribution of alpha particles in the ternary fission of $^{244-248}$Cm isotopes, but its mass distribution has not been studied so far. We have computed the emission probability of long range alpha particle emitted in the ternary fission of $^{242-248}$Cm isotopes and are in good agreement with the experimental data [23, 24]. Using our formalism we have calculated the mass distribution of heavy fragments in the ternary fission of $^{244-252}$Cm isotopes and have predicted the fragments with highest yield. Our study shows that the fragments $^{109}$Tc, $^{131}$Sb, $^{116}$Pd and $^{124}$Cd from $^{244}$Cm; $^{109}$Tc, $^{114}$Ru, $^{116}$Pd, $^{126}$Cd, $^{128}$Sn and $^{133}$Sb from $^{246}$Cm; $^{114}$Ru, $^{116}$Pd, $^{128}$Cd and $^{130}$Sn from $^{248}$Cm; $^{114}$Ru and $^{132}$Sn from $^{250}$Cm; and $^{115}$Ru, $^{116}$Ru, $^{132}$Sn and $^{133}$Sn from $^{252}$Cm have relative yield greater than 10%. We hope that our prediction on the yield of heavy fragments in ternary fission of $^{244-252}$Cm isotopes will guide the future experiments and hope these fragments can be detected using triple gamma coincidence method with the help of Gammasphere as done in the case of alpha accompanied ternary fission of $^{252}$Cf isotope [21].

## 4. Summary

With $^4$He as light charged particle, the relative yield is calculated by taking the interacting barrier as the sum of Coulomb and proximity potential with fragments in equatorial configuration for the ternary fission of $^{244}$Cm, $^{246}$Cm, $^{248}$Cm, $^{250}$Cm and $^{252}$Cm. In the ternary fission of $^{244}$Cm, the highest yield is found for the splitting $^{110}$Ru+$^{4}$He+$^{130}$Sn which possess nearly doubly magic nuclei $^{130}$Sn. The highest yield found for alpha accompanied the ternary fragmentation of $^{246}$Cm, $^{248}$Cm, $^{250}$Cm and $^{252}$Cm is with $^{110}$Ru+$^{4}$He+$^{132}$Sn, $^{112}$Ru+$^{4}$He+$^{132}$Sn, $^{114}$Ru+$^{4}$He+$^{132}$Sn and $^{116}$Ru+$^{4}$He+$^{132}$Sn respectively, all of which possesses a higher $Q$ value and doubly magic nuclei $^{132}$Sn. Hence for the most favorable fragment combination to occur in ternary fission, the presence of doubly magic nuclei and high $Q$ values play a crucial role. The emission probabilities and kinetic energies of long range alpha particle are computed for the isotopes $^{242}$Cm, $^{244}$Cm, $^{246}$Cm, $^{248}$Cm and are found to be in good agreement with the experimental data. The yield obtained for the equatorial configuration is higher than that of the collinear configuration. It is also found that the relative yield for binary exit channel is found to be higher than that of ternary fragmentation (both equatorial and collinear configuration). The studies on the influence of deformation in the alpha accompanied ternary fission of $^{244-252}$Cm isotopes reveal that the ground state deformation has an important role in ternary fission as that of shell effect.

**Acknowledgments**

The author KPS would like to thank the University Grants Commission, Govt. of India for the financial support under Major Research Project. No.42-760/2013 (SR) dated 22-03-2013.

**References**

[1] Alvarez L W, as reported by Farwell G, Segre E and Wiegand C 1947 *Phys. Rev.* **71** 327

[2] Cassels J M, Dainty J, Feather N and Green L L 1947 *Proc. Roy. Soc. A* **191** 428

[3] Malkin L Z, Alkhazov I D, Krisvokhatskii A S, Petrzhak K A and Belov L M 1964 *Translated from Atomnaya Energiya* **16** 148

[4] Sandulescu A, Florescu A, Carstoiu F and Greiner W 1997 *J. Phys. G: Nucl. Part. Phys.* **23** L7

[5] Sandulescu A, Carstoiu F, Misicu S, Florescu A, Ramayya A V, Hamilton J H and Greiner W 1998 *J. Phys. G: Nucl. Part. Phys.* **24** 181

[6] Sandulescu A, Carstoiu F, Bulboaca I and Greiner W 1999 *Phys. Rev. C* **60** 044613

[7] Royer G, Degiorgio K, Dubillot M and Leonard E 2008 *J. Phys. Conference Series* **111** 012052

[8] Royer G and Mignen J 1992 *J. Phys. G: Nucl. Part. Phys.* **18** 1781

[9] Royer G, Haddad F and Mignen J 1992 *J. Phys. G: Nucl. Part. Phys.* **18** 2015

[10] Delion D S, Sandulescu A and Greiner W 2002 *J. Phys. G: Nucl. Part. Phys.* **28** 2921

[11] Delion D S, Sandulescu A and Greiner W 2003 *J. Phys. G: Nucl. Part. Phys.* **29** 317

[12] Delion D S, Florescu A, and Sandulescu A 2001 *Phys. Rev. C* **63** 044312

[13] Gherghescu R A and Carjan N 2009 *J. Phys. G: Nucl. Part. Phys.* **36** 025106

[14] Jandel M, Kliman J, Krupa L, Morha M, Hamilton J H, Kormicki J, Ramayya A V, Hwang J K, Luo Y X, Fong D, Gore P, Akopian G M, Oganessian Yu Ts, Rodin A M, Fomichev A S, Popeko G S, Daniel A V, Rasmssen J O, Macchiavelli A O, Stoyer M A, Donangelo R and Cole J D 2002 *J. Phys. G: Nucl. Part. Phys.* **28** 2893

[15] Andreev A V, Adamian G G, Antonenko N V, Ivanova S P, Kuklin S N and Scheid W 2006 *Eur. Phys. J. A* **30** 579

[16] Mirzaei V and Miri-Hakimabad H 2012 *Rom. Rep. Phys.* **64** 50

[17] Florescu A, Sandulescu A, Delion D S, Hamilton J H, Ramayya A V and Greiner W 2000 *Phys. Rev. C* **61** 051602

[18] Misicu S, Hess P O and Greiner W 2001 *Phys. Rev.C* **63** 054308

[19] Ronen Y 2002 *Ann. Nucl. Energy* **29** 1013

[20] Hamilton J H, Ramayya A V, Hwang J K, Kormicki J, Babu B R S, Sandulescu A, Florescu A, Greiner W, Ter-Akopian G M, Oganessian Yu Ts, Daniel A V, Zhu S J, Wang M G, Ginter T, Deng J K, Ma W C, Popeko G S, Lu Q H, Jones E, Dodder R, Gore P, Nazarewicz W, Rasmussen J O, Asztalos S, Lee I Y, Chu S Y, Gregorich K E, Macchiavell A O, Mohar M F, Prussino S, Stoyero M A, Lougheedo R W, Moody K J, Wild J F, Bernstein L A, Becker J A, Cole J D, Aryaeinejad R, Dardenne Y X, Drigert M W, Butler-Moore K, Donangel R and Griffin H C 1997 *Prog. Part. Nucl. Phys.* **38** 273

[21] Ramayya A V, Hamilton J H, Hwang J K, Peker L K, Kormicki J, Babu B R S, Ginter T N, Sandulescu A, Florescu A, Carstoiu F, Greiner W, Ter-Akopian G M, Oganessian Yu Ts, Daniel A V, Ma W C, Varmette P G, Rasmussen J O, Asztalos S J, Chu S Y, Gregorich K E, Macchiavelli A O, Macleod R W, Cole J D, Aryaeinejad R, Butler-Moore K, Drigert M W, Stoyer M A, Bernstein L A, Lougheed R W, Moody K J, Prussin S G, Zhu S J, Griffin H C and Donangelo R 1998 *Phys. Rev. C* **57** 2370

[22] Ramayya A V, Hamilton J H, Hwang J K 2007 *Rom. Rep. Phys.* **59**, 595

[23] Serot O, Wagemans C 1998 *Nucl. Phys. A* **641** 34

[24] Vermote S, Wagemans C, Serot O, Heyse J, Van Gils J, Soldner T and Geltenbort P 2008 *Nucl. Phys. A* **806** 1

[25] Vermote S, Wagemans C, Serot O, Heyse J, Van Gils J, Soldner T, Geltenbort P, AlMahamid I, Tian G, Rao L 2010 *Nucl. Phys. A* **837** 176

[26] Manimaran K and Balasubramaniam M 2009 *Phys. Rev. C* **79** 024610

[27] Manimaran K and Balasubramaniam M 2010 *Eur. Phys. J. A* **45** 293

[28] Manimaran K and Balasubramaniam M 2010 *J. Phys. G: Nucl. Part. Phys.* **37** 045104

[29] Manimaran K and Balasubramaniam M 2011 *Phys. Rev. C* **83**, 034609

[30] Thakur S, Kumar R, Vijayaraghavan K R and Balasubramaniam M 2013 *Int. J. Mod. Phy. E* **22** 1350014

[31] Santhosh K P, Sreejith Krishnan, Priyanka B 2014 *Eur. Phys. J. A* **50**, 66

[32] Oertzen von W, Pyatkov Y V, Kamanin D 2013 *Acta. Phys. Pol. B* **44**, 447

[33] Santhosh K P, Jayesh G J, Sabina S 2010 *Phys. Rev. C* **82** 064605

[34] Santhosh K P, Priyanka B 2013 *Phys. Rev. C* **87** 064611

[35] Santhosh K P, Biju R K, Sabina S and Joseph A 2008 *Phys. Scr.* **77** 065201

[36] Santhosh K P and Bobby Jose V 2014 *Nucl. Phys. A* **922** 191

[37] NuDat2.5, http://www.nndc.bnl.gov

[38] Ogloblin A A, Bonetti R, Denisov V A, Guglielmetti A, Itkis M G, Mazzocchi C, Mikheev V L, Oganessian Yu Ts, Pik-Pichak G A, Poli G, Pirozhkov S M, Semochkin V M, Shigin V A, Shvetsov I K, and Tretyakova S P 2000 Phys. Rev. **C 61** 034301

[39] Santhosh K P, Priyanka B and Unnikrishnan M S 2012 Nucl. Phys. **A 889** 29

[40] Blocki J, Randrup J, Swiatecki W J, Tsang C F 1977 *Ann. Phys. (N.Y)* **105** 427

[41] Blocki J, Swiatecki W J 1981 *Ann. Phys. (N.Y)* **132** 53

[42] Shi Y J and Swiatecki W J 1985 *Nucl. Phys. A* **438** 450

[43] Malik S S and Gupta R K 1989 *Phys.Rev.C* **39** 1992

[44] Myers W D and Swiatecki W J 1967 *Ark. Fys.* 36 343

[45] Gupta R K, in *Heavy elements and related new phenomena* edited by Gupta R K and Greiner W (World Scientific Pub., Singapore, 1999) vol II, p. 730

[46] Wang M, Audi G, Wapstra A H, Kondev F G, MacCormick M, Xu X and Pfeiffer B 2012 *Chin. Phys. C* **36** 1603

[47] Royer G, Haddad F and Mignen J 1992 J. Phys. G: Nucl. Part. Phys. **18** 2015

[48] Carjan N 1976 *J. de Phys.* **37** 1279

[49] Blendowske R, Fliessbach T and Walliser H 1991 *Z. Phys. A* **339** 121

[50] Fraenkel Z 1967 *Phys. Rev.* **156** 1283

[51] Viola V E, Kwiatkowski K and Walker M 1985 *Phys. Rev. C* **31** 1550

[52] Herbach C M, Hilscher D, Tishchenko V G, Gippner P, Kamanin D V, Oertzen W von, Ortlepp H G, Penionzhkevich Yu E, Pyatkov Yu V, Renz G, Schilling K D, Strekalovsky O V, Wagner W, Zhuchko V E 2002 *Nucl. Phys. A* **712** 207

[53] Perfilov N A, Soloveva Z I, and Filov R A 1964 *J. Exptl. Theoret. Phys.* **19** 1515

[54] Wong C Y 1973 Phys. Rev. Lett. **31** 766

[55] Malhotra N and Gupta R K 1985 Phys. Rev. **C 31** 1179

[56] Gupta R K, Balasubramaniam M, Kumar R, Singh N, Manhas M and Greiner W 2005 J. Phys. G: Nucl. Part. Phys. **31** 631

[57] Baltz A J and Bayman B F 1982 Phys. Rev. **C 26** 1969

[58] http://w.w.w.nds.iaea.org/RIPL-2

[59] Moller P, Nix J R and Kratz K L 1997 At. Data Nucl. Data Tables **66** 131

**Table 1.** The calculated and experimental emission probability [23, 24] for the ternary α's of different curium isotopes. The computed spectroscopic factor $S_\alpha$ and $P_{LRA}$ are also listed.

| Isotope | $S_\alpha$ | $P_{LRA}$ | $\frac{LRA}{B}[10^{-3}]$ | $\left(\frac{LRA}{B}\right)_{EXP.}[10^{-3}]$ |
|---|---|---|---|---|
| $^{242}$Cm | 0.0249 | 0.1141 | 2.84 | 3.34 ± 0.26 |
| $^{244}$Cm | 0.0243 | 0.1128 | 2.74 | 2.73 ± 0.20 |
| $^{246}$Cm | 0.0247 | 0.1614 | 3.98 | 2.49 ± 0.12 |
| $^{248}$Cm | 0.0271 | 0.1713 | 4.64 | 2.30 ± 0.30 |

**Table 2.** Comparison of the calculated kinetic energy of alpha particle $E_\alpha$ emitted in the ternary fission of $^{242-252}$Cm isotopes with the experimental data [24, 53].

| Fragmentation channel | $E_\alpha$ (MeV) Calc. | $E_\alpha$ (MeV) Expt. | Fragmentation channel | $E_\alpha$ (MeV) Calc. | $E_\alpha$ (MeV) Expt. |
|---|---|---|---|---|---|
| $^{242}$Cm → $^{96}$Sr + $^{4}$He + $^{142}$Ba | 15.14 | | $^{248}$Cm → $^{100}$Sr + $^{4}$He + $^{144}$Ba | 15.53 | |
| $^{242}$Cm → $^{98}$Zr + $^{4}$He + $^{140}$Xe | 15.31 | | $^{248}$Cm → $^{102}$Sr + $^{4}$He + $^{142}$Ba | 15.68 | |
| $^{242}$Cm → $^{100}$Zr + $^{4}$He + $^{138}$Xe | 15.46 | | $^{248}$Cm → $^{104}$Zr + $^{4}$He + $^{140}$Xe | 15.82 | |
| $^{242}$Cm → $^{102}$Zr + $^{4}$He + $^{136}$Xe | 15.60 | | $^{248}$Cm → $^{106}$Zr + $^{4}$He + $^{138}$Xe | 15.95 | |
| $^{242}$Cm → $^{104}$Mo + $^{4}$He + $^{134}$Te | 15.72 | | $^{248}$Cm → $^{108}$Mo + $^{4}$He + $^{136}$Te | 16.06 | |
| $^{242}$Cm → $^{106}$Mo + $^{4}$He + $^{132}$Te | 15.83 | 15.50 ± 1.00 | $^{248}$Cm → $^{110}$Mo + $^{4}$He + $^{134}$Te | 16.16 | 15.97 ± 0.12 |
| $^{242}$Cm → $^{108}$Ru + $^{4}$He + $^{130}$Sn | 15.92 | | $^{248}$Cm → $^{112}$Ru + $^{4}$He + $^{132}$Sn | 16.24 | |
| $^{242}$Cm → $^{110}$Ru + $^{4}$He + $^{128}$Sn | 16.00 | | $^{248}$Cm → $^{114}$Ru + $^{4}$He + $^{130}$Sn | 16.31 | |
| $^{242}$Cm → $^{112}$Ru + $^{4}$He + $^{126}$Sn | 16.06 | | $^{248}$Cm → $^{116}$Ru + $^{4}$He + $^{128}$Sn | 16.36 | |
| $^{242}$Cm → $^{114}$Pd + $^{4}$He + $^{124}$Cd | 16.11 | | $^{248}$Cm → $^{118}$Pd + $^{4}$He + $^{126}$Cd | 16.40 | |
| $^{242}$Cm → $^{116}$Pd + $^{4}$He + $^{122}$Cd | 16.14 | | $^{248}$Cm → $^{120}$Pd + $^{4}$He + $^{124}$Cd | 16.42 | |
| $^{242}$Cm → $^{118}$Pd + $^{4}$He + $^{120}$Cd | 16.16 | | $^{248}$Cm → $^{122}$Pd + $^{4}$He + $^{122}$Cd | 16.43 | |
| $^{244}$Cm → $^{96}$Sr + $^{4}$He + $^{144}$Ba | 15.16 | | $^{250}$Cm → $^{98}$Sr + $^{4}$He + $^{148}$Ba | 15.38 | |
| $^{244}$Cm → $^{98}$Sr + $^{4}$He + $^{142}$Ba | 15.33 | | $^{250}$Cm → $^{100}$Sr + $^{4}$He + $^{146}$Ba | 15.55 | |
| $^{244}$Cm → $^{100}$Zr + $^{4}$He + $^{140}$Xe | 15.49 | | $^{250}$Cm → $^{102}$Zr + $^{4}$He + $^{144}$Xe | 15.71 | |
| $^{244}$Cm → $^{102}$Zr + $^{4}$He + $^{138}$Xe | 15.63 | | $^{250}$Cm → $^{104}$Zr + $^{4}$He + $^{142}$Xe | 15.85 | |
| $^{244}$Cm → $^{104}$Zr + $^{4}$He + $^{136}$Xe | 15.76 | | $^{250}$Cm → $^{106}$Zr + $^{4}$He + $^{140}$Xe | 15.98 | |
| $^{244}$Cm → $^{106}$Mo + $^{4}$He + $^{134}$Te | 15.88 | | $^{250}$Cm → $^{108}$Mo + $^{4}$He + $^{138}$Te | 16.10 | |
| $^{244}$Cm → $^{108}$Ru + $^{4}$He + $^{132}$Sn | 15.97 | 15.99 ± 0.08 | $^{250}$Cm → $^{110}$Mo + $^{4}$He + $^{136}$Te | 16.21 | |
| $^{244}$Cm → $^{110}$Ru + $^{4}$He + $^{130}$Sn | 16.06 | | $^{250}$Cm → $^{112}$Mo + $^{4}$He + $^{134}$Te | 16.29 | |
| $^{244}$Cm → $^{112}$Ru + $^{4}$He + $^{128}$Sn | 16.12 | | $^{250}$Cm → $^{114}$Ru + $^{4}$He + $^{132}$Sn | 16.37 | |
| $^{244}$Cm → $^{114}$Ru + $^{4}$He + $^{126}$Sn | 16.18 | | $^{250}$Cm → $^{116}$Ru + $^{4}$He + $^{130}$Sn | 16.43 | |
| $^{244}$Cm → $^{116}$Pd + $^{4}$He + $^{124}$Cd | 16.22 | | $^{250}$Cm → $^{118}$Pd + $^{4}$He + $^{128}$Cd | 16.47 | |
| $^{244}$Cm → $^{118}$Pd + $^{4}$He + $^{122}$Cd | 16.24 | | $^{250}$Cm → $^{120}$Pd + $^{4}$He + $^{126}$Cd | 16.50 | |
| $^{244}$Cm → $^{120}$Pd + $^{4}$He + $^{120}$Cd | 16.25 | | $^{250}$Cm → $^{122}$Pd + $^{4}$He + $^{124}$Cd | 16.52 | |
| $^{246}$Cm → $^{96}$Sr + $^{4}$He + $^{146}$Ba | 15.17 | | $^{252}$Cm → $^{100}$Sr + $^{4}$He + $^{148}$Ba | 15.56 | |
| $^{246}$Cm → $^{98}$Sr + $^{4}$He + $^{144}$Ba | 15.35 | | $^{252}$Cm → $^{102}$Sr + $^{4}$He + $^{146}$Ba | 15.73 | |
| $^{246}$Cm → $^{100}$Sr + $^{4}$He + $^{142}$Ba | 15.51 | | $^{252}$Cm → $^{104}$Zr + $^{4}$He + $^{144}$Xe | 15.88 | |
| $^{246}$Cm → $^{102}$Zr + $^{4}$He + $^{140}$Xe | 15.66 | | $^{252}$Cm → $^{106}$Zr + $^{4}$He + $^{142}$Xe | 16.02 | |
| $^{246}$Cm → $^{104}$Zr + $^{4}$He + $^{138}$Xe | 15.79 | | $^{252}$Cm → $^{108}$Zr + $^{4}$He + $^{140}$Xe | 16.14 | |
| $^{246}$Cm → $^{106}$Mo + $^{4}$He + $^{136}$Te | 15.92 | | $^{252}$Cm → $^{110}$Mo + $^{4}$He + $^{138}$Te | 16.25 | |
| $^{246}$Cm → $^{108}$Mo + $^{4}$He + $^{134}$Te | 16.02 | 16.41 ± 0.20 | $^{252}$Cm → $^{112}$Mo + $^{4}$He + $^{136}$Te | 16.34 | |
| $^{246}$Cm → $^{110}$Ru + $^{4}$He + $^{132}$Sn | 16.11 | | $^{252}$Cm → $^{114}$Ru + $^{4}$He + $^{134}$Sn | 16.42 | |
| $^{246}$Cm → $^{112}$Ru + $^{4}$He + $^{130}$Sn | 16.18 | | $^{252}$Cm → $^{116}$Ru + $^{4}$He + $^{132}$Sn | 16.49 | |
| $^{246}$Cm → $^{114}$Ru + $^{4}$He + $^{128}$Sn | 16.24 | | $^{252}$Cm → $^{118}$Ru + $^{4}$He + $^{130}$Sn | 16.54 | |
| $^{246}$Cm → $^{116}$Pd + $^{4}$He + $^{126}$Cd | 16.29 | | $^{252}$Cm → $^{120}$Pd + $^{4}$He + $^{128}$Cd | 16.58 | |
| $^{246}$Cm → $^{118}$Pd + $^{4}$He + $^{124}$Cd | 16.32 | | $^{252}$Cm → $^{122}$Pd + $^{4}$He + $^{126}$Cd | 16.60 | |
| $^{246}$Cm → $^{120}$Pd + $^{4}$He + $^{122}$Cd | 16.34 | | $^{252}$Cm → $^{124}$Pd + $^{4}$He + $^{124}$Cd | 16.61 | |

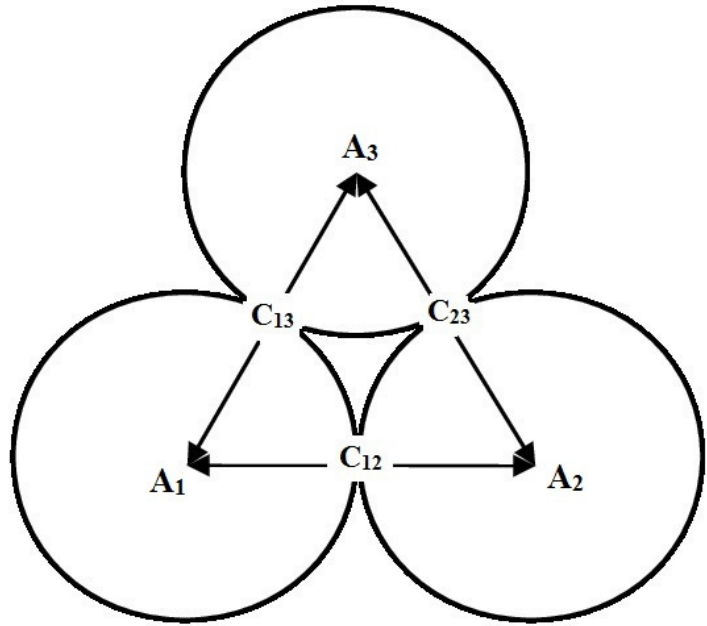
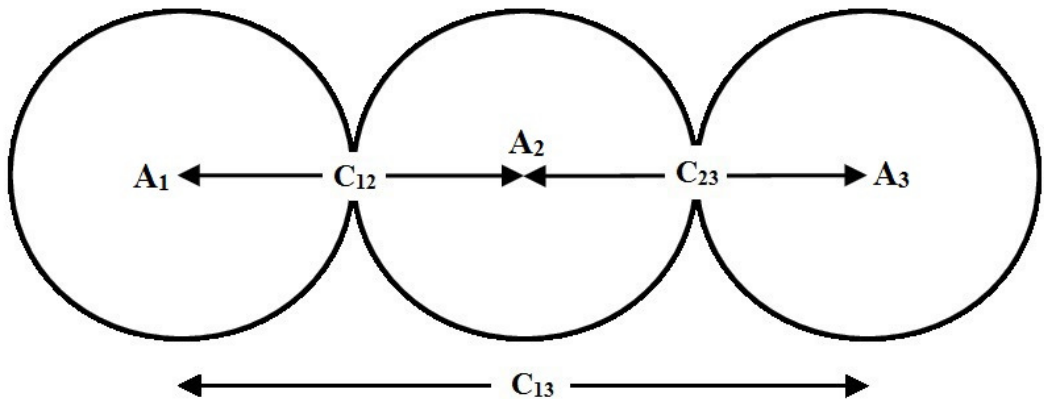

**Figure 1.** The touching configuration of three spherical fragments in a) equatorial configuration b) collinear configuration.

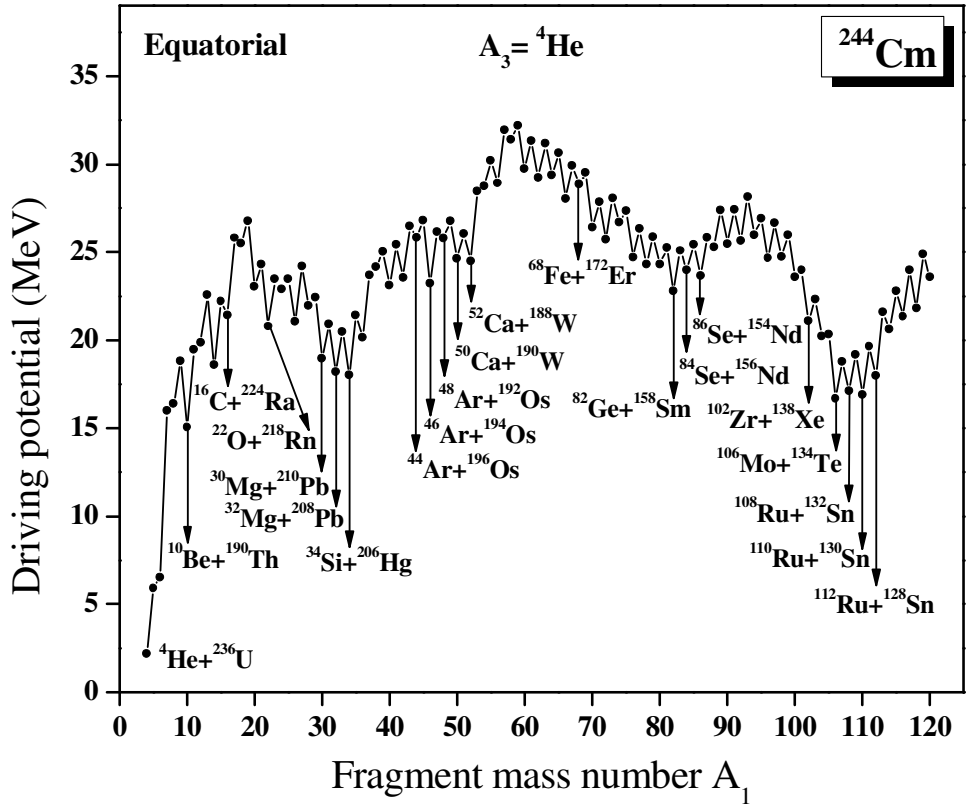

**Figure 2.** The driving potential for $^{244}$Cm isotope with $^4$He as light charged particle, with fragments in the equatorial configuration plotted as a function of mass number $A_1$.

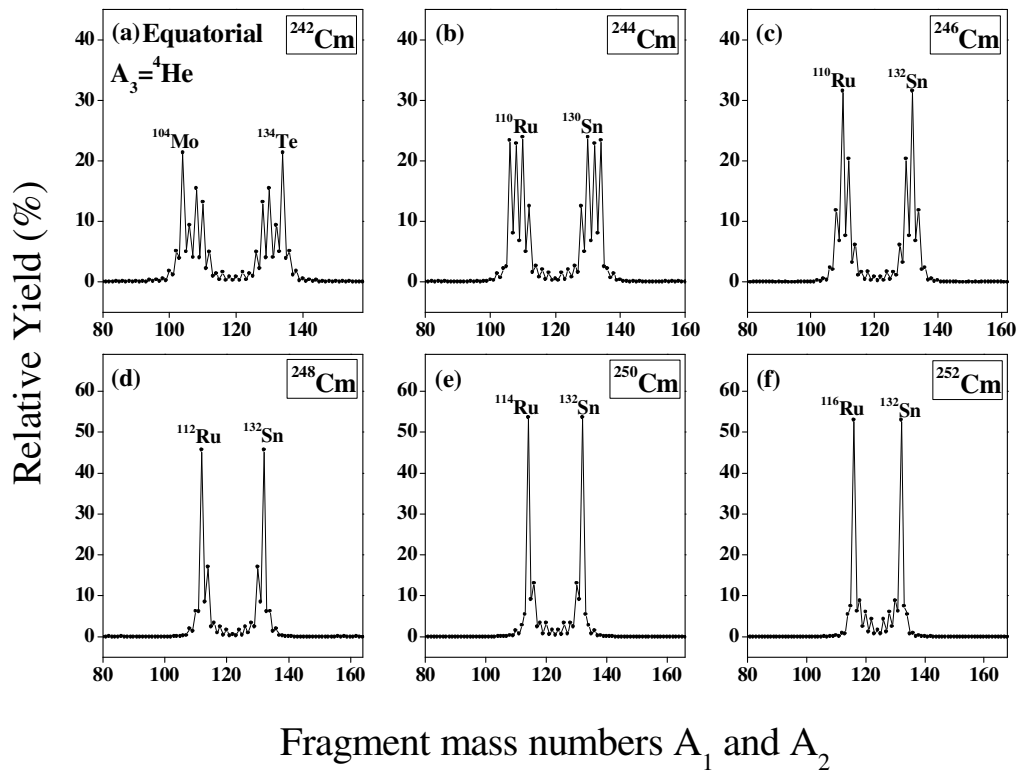

**Figure 3.** The calculated yields for the charge minimized third fragment $^{4}$He in the case of equatorial configuration plotted as a function of mass numbers $A_1$ and $A_2$ for the isotopes $^{242-252}$Cm. The fragment combinations with highest yield are labeled.

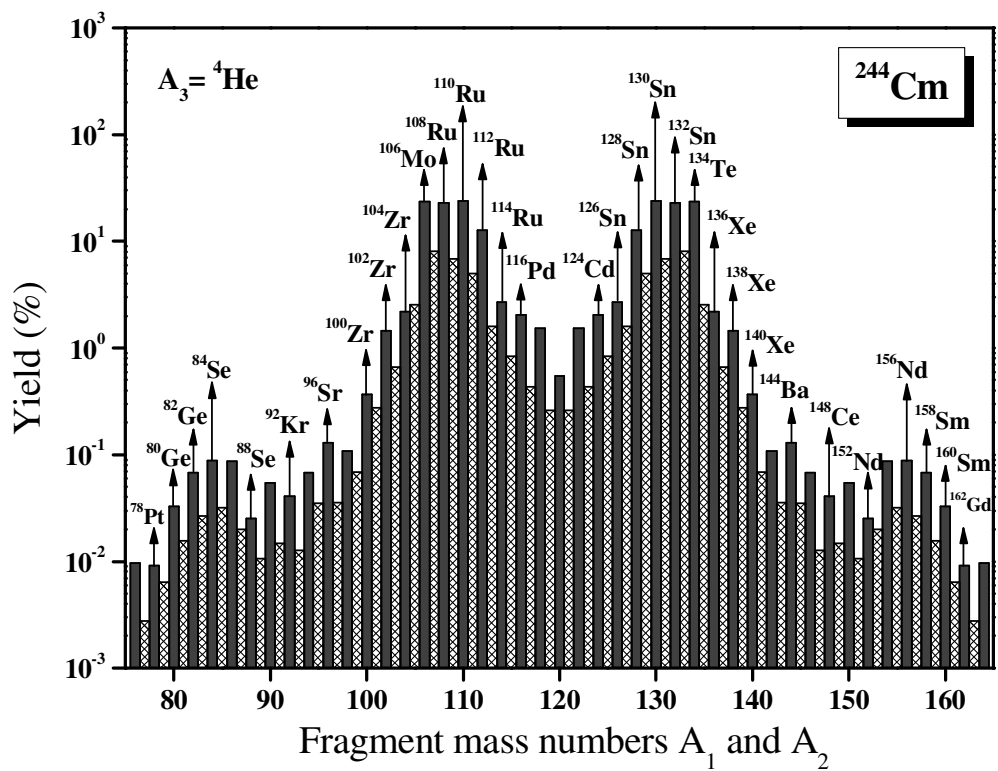

**Figure 4.** The calculated yields for the charge minimized third fragment $^4$He in the case of equatorial configuration for $^{244}$Cm plotted as a function of mass numbers $A_1$ and $A_2$.

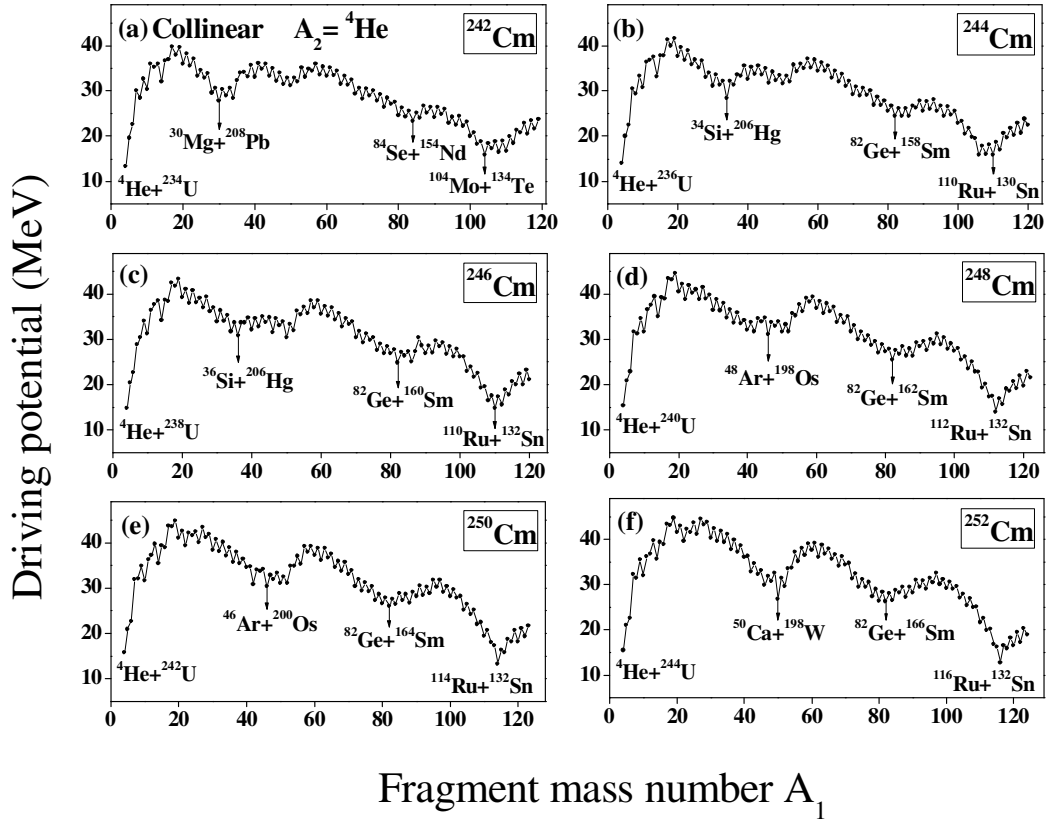

**Figure 5.** The driving potential for $^{242-252}$Cm isotope with $^4$He as light charged particle in the case of collinear configuration plotted as a function of mass number $A_1$.

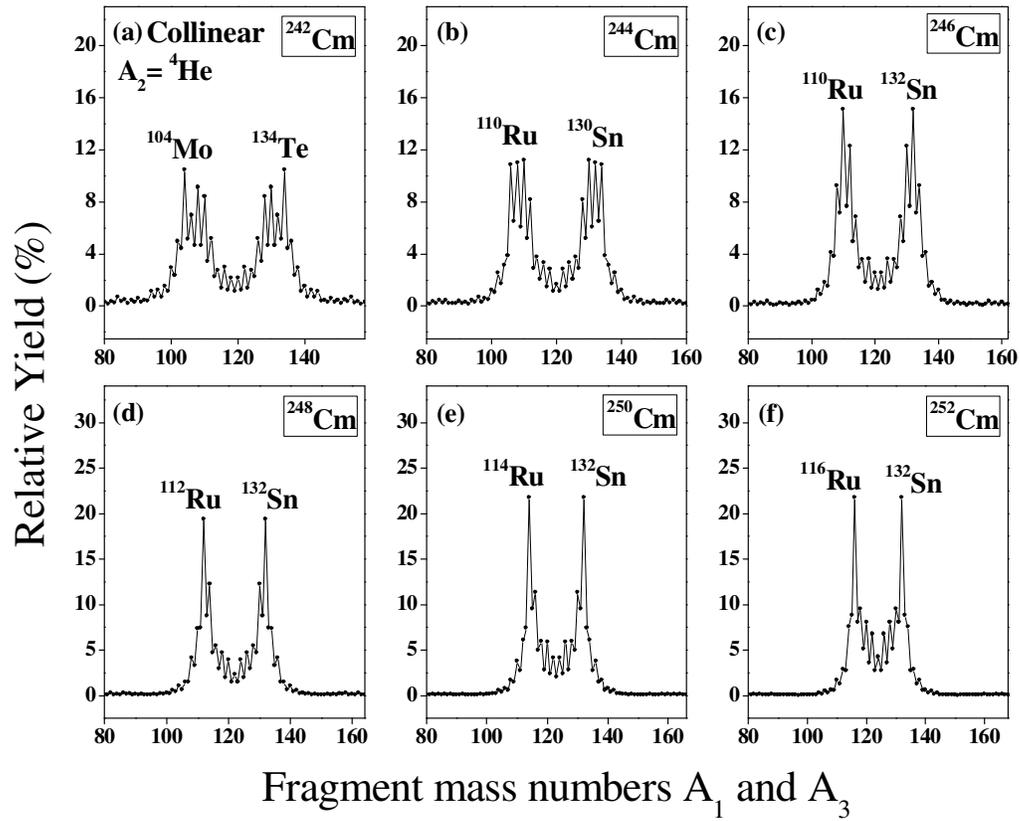

**Figure 6.** The calculated yields for the ternary fission of $^{242-252}$Cm isotopes with charge minimized third fragment $^4$He in the case of collinear configuration plotted as a function of mass numbers $A_1$ and $A_3$. The fragment combinations with highest yield are labeled.

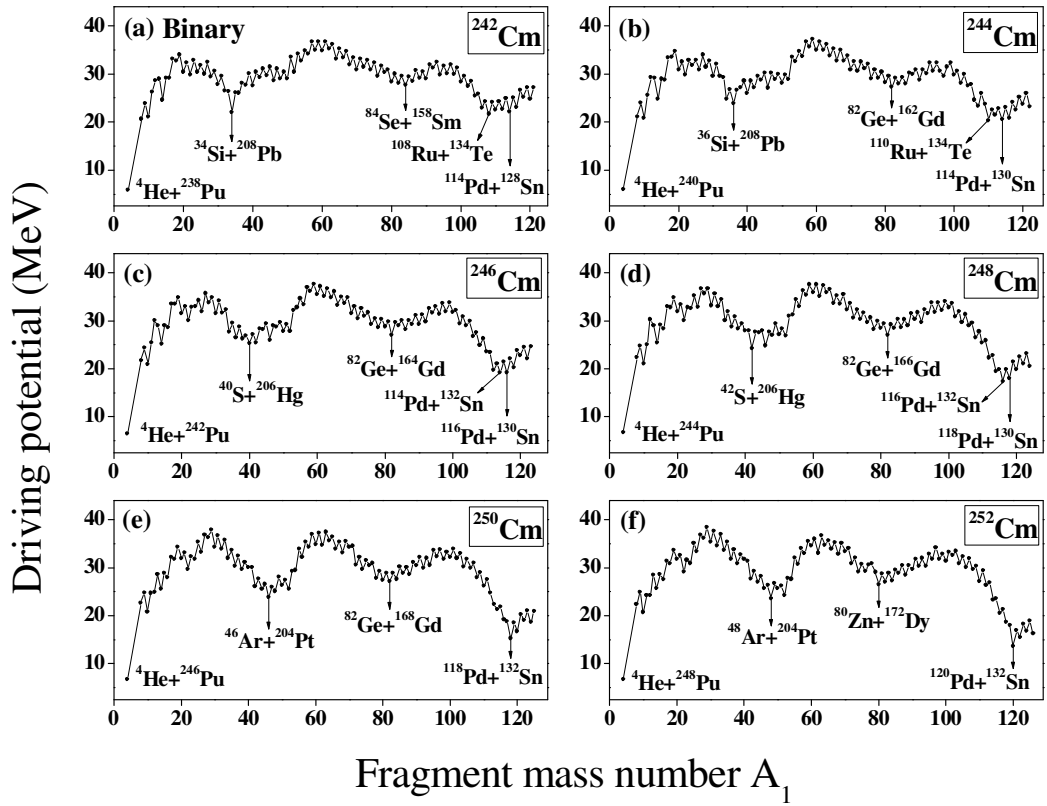

**Figure 7.** The driving potential for the binary fission of $^{242-252}$Cm isotope plotted as a function of mass number $A_1$.

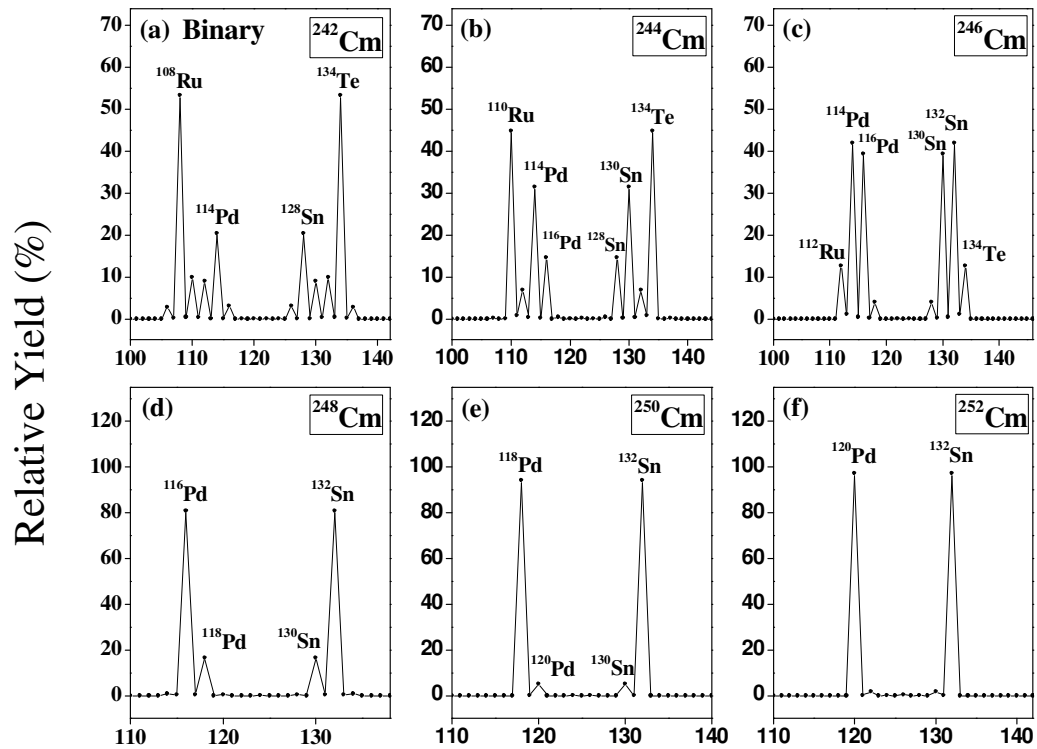

**Figure 8.** The calculated relative yields for the binary fission of $^{242-252}$Cm isotopes plotted as a function of mass numbers $A_1$ and $A_2$. The fragment combinations with highest yield are labeled.

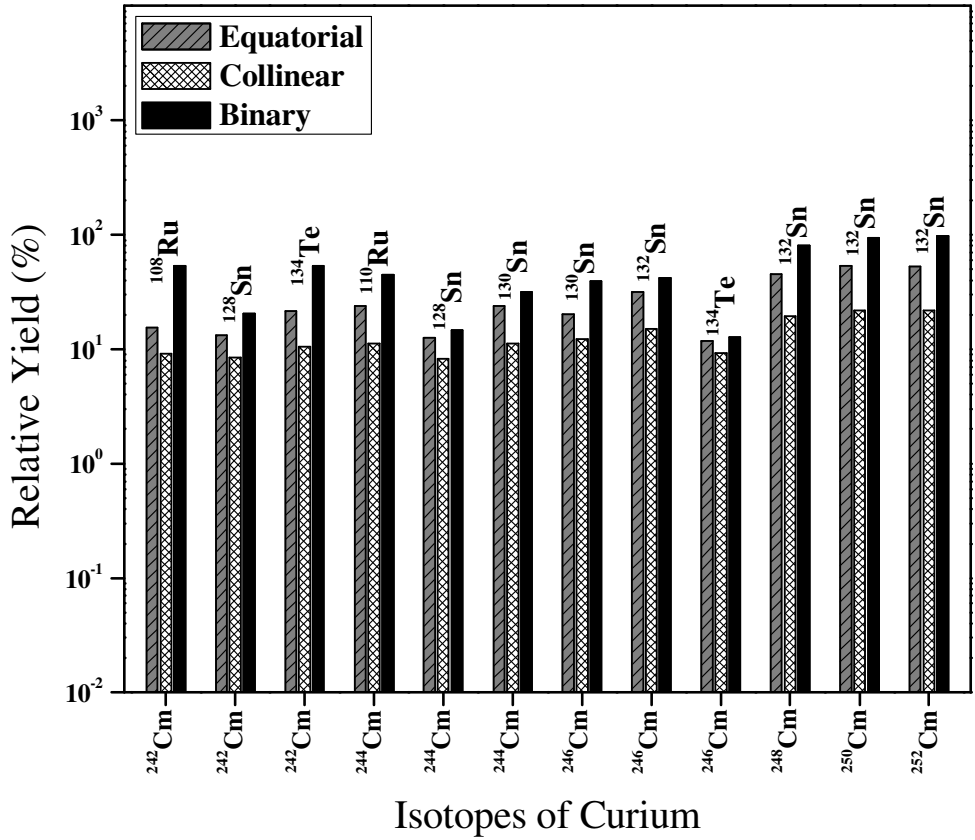

**Figure 9.** Comparison of relative yields for the $^4$He accompanied ternary fission (both equatorial and collinear configurations) of $^{242-252}$Cm with the yield for binary fission.

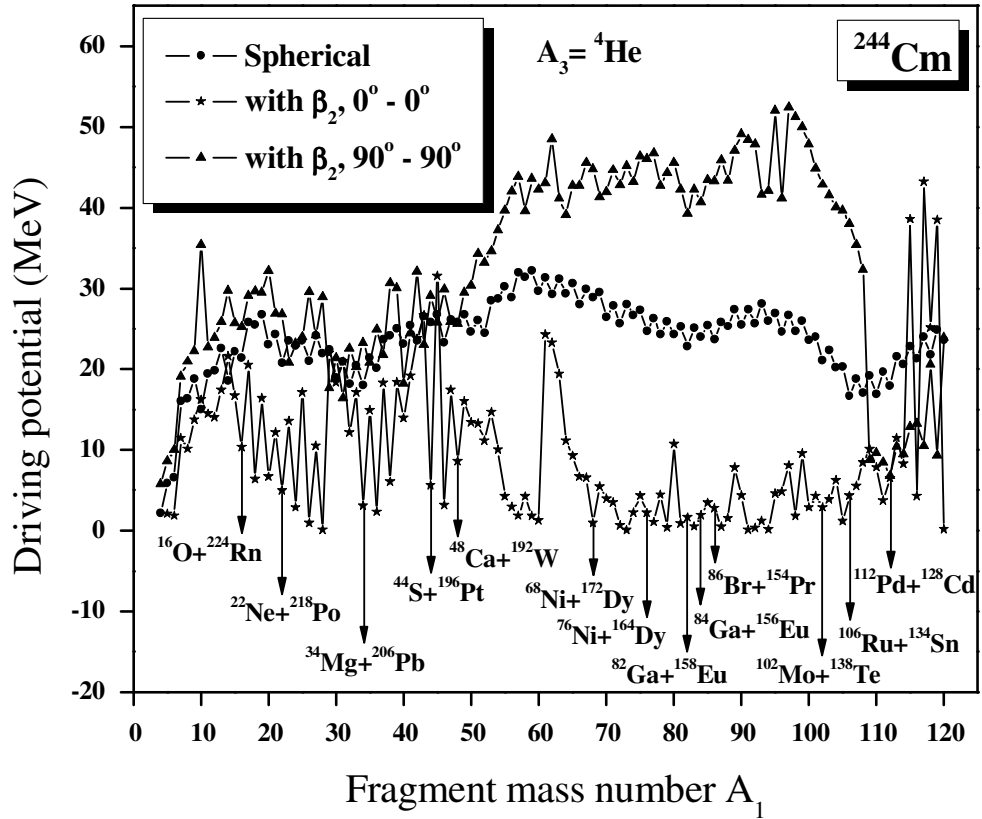

**Figure 10.** The driving potential for $^{244}$Cm isotope with $^4$He as light charged particle with the inclusion of quadrupole deformation $β_2$ and for different orientation plotted as a function of mass number $A_1$.

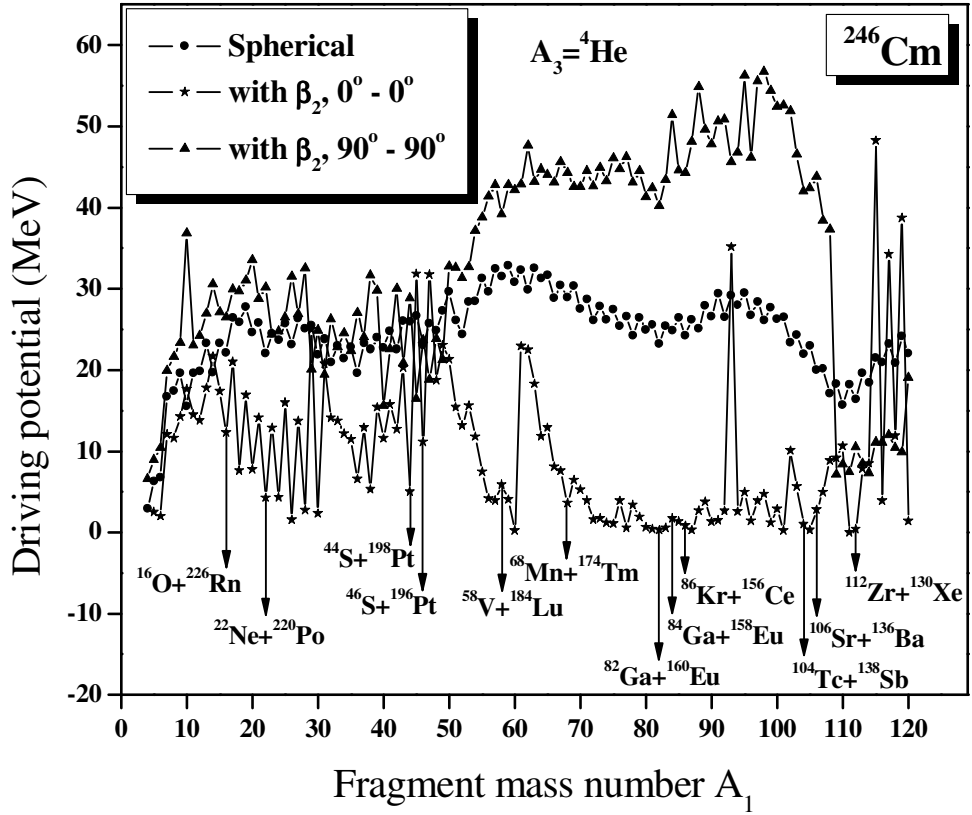

**Figure 11.** The driving potential for $^{246}$Cm isotope with $^4$He as light charged particle with the inclusion of quadrupole deformation $\beta_2$ and for different orientation plotted as a function of mass number $A_1$.

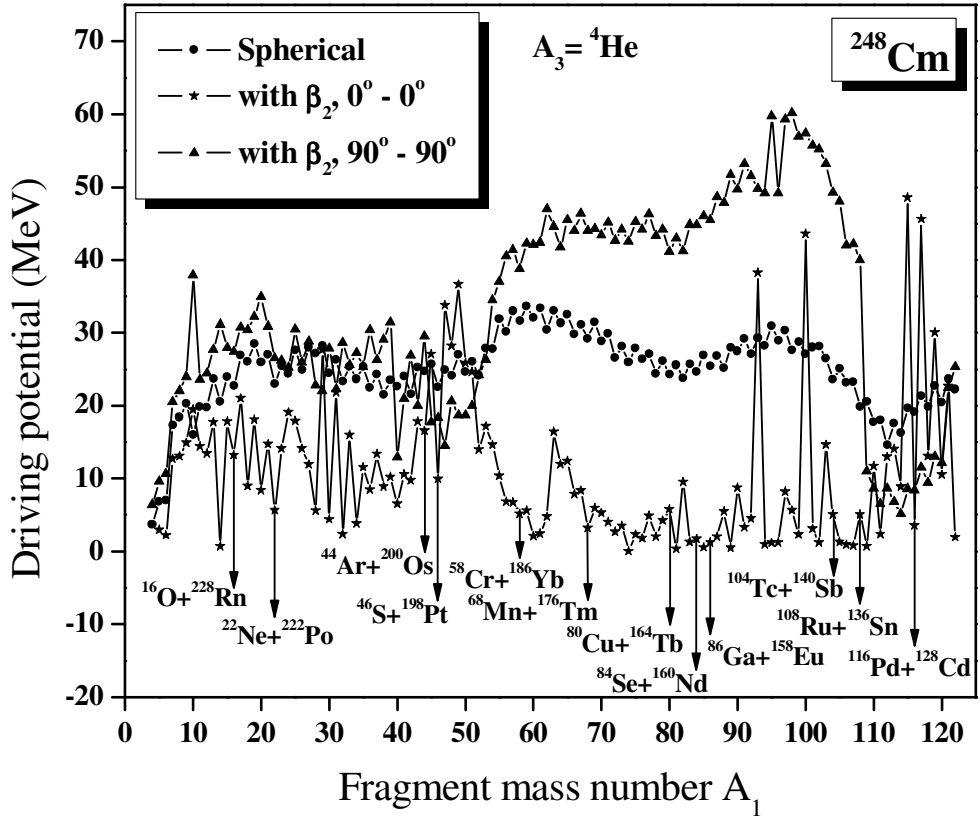

**Figure 12.** The driving potential for $^{248}$Cm isotope with $^4$He as light charged particle with the inclusion of quadrupole deformation $\beta_2$ and for different orientation plotted as a function of mass number $A_1$.

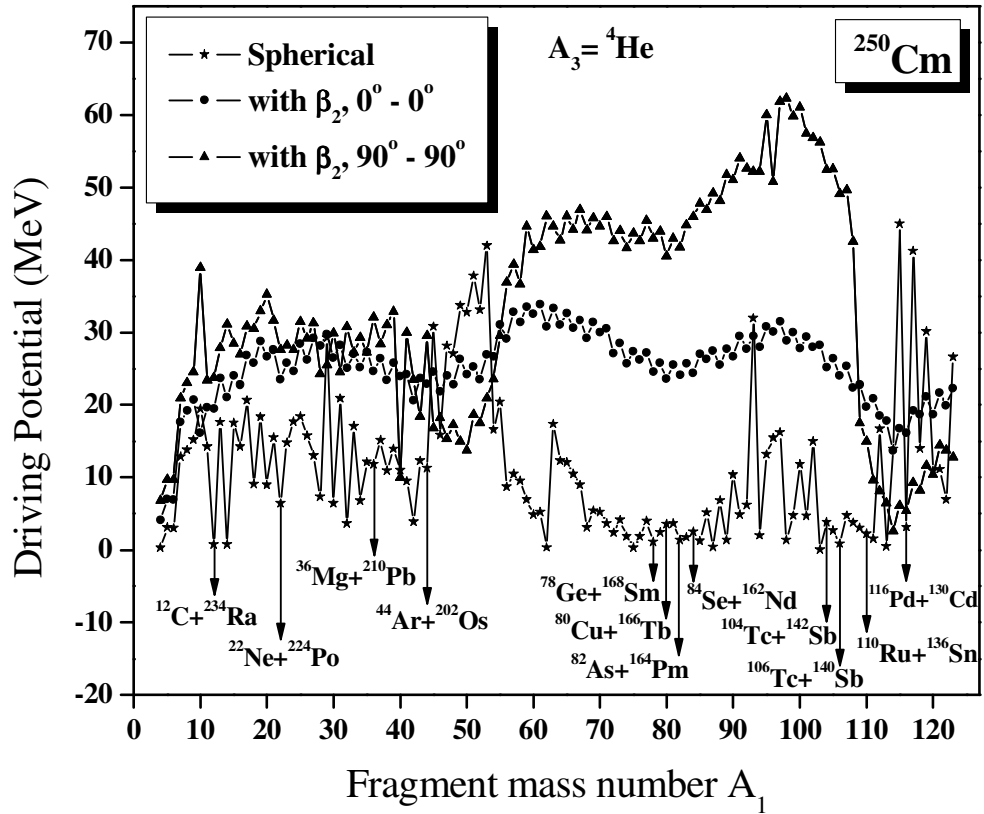

**Figure 13.** The driving potential for $^{250}$Cm isotope with $^4$He as light charged particle with the inclusion of quadrupole deformation $\beta_2$ and for different orientation plotted as a function of mass number $A_1$.

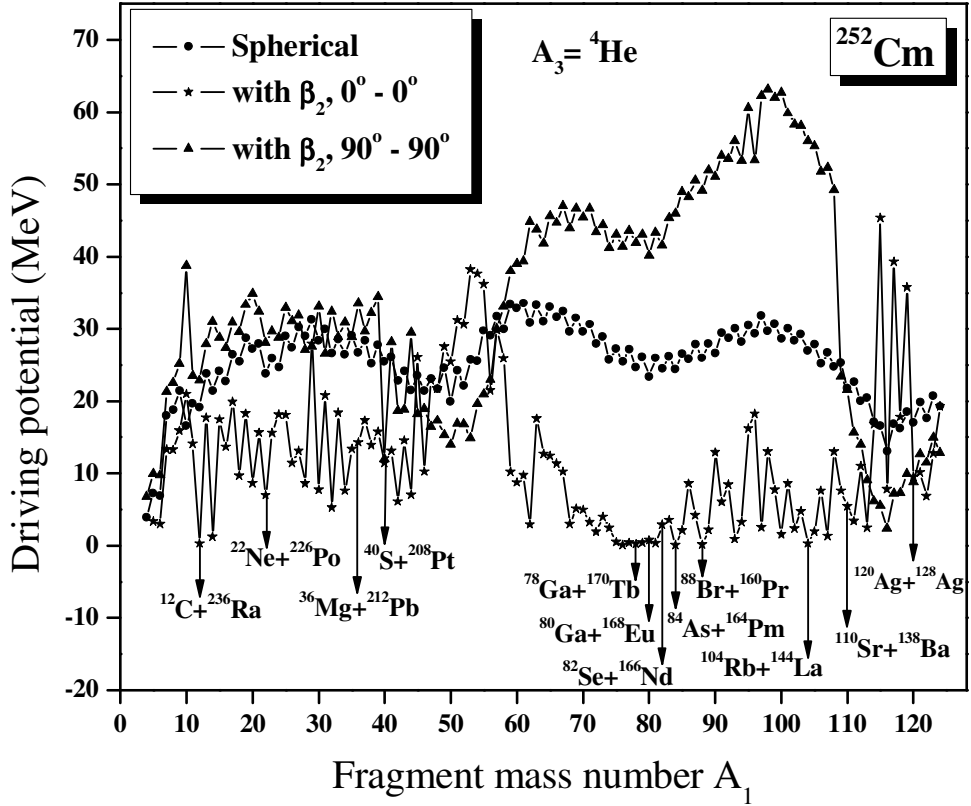

**Figure 14.** The driving potential for $^{252}$Cm isotope with $^{4}$He as light charged particle with the inclusion of quadrupole deformation $\beta_2$ and for different orientation plotted as a function of mass number $A_1$.

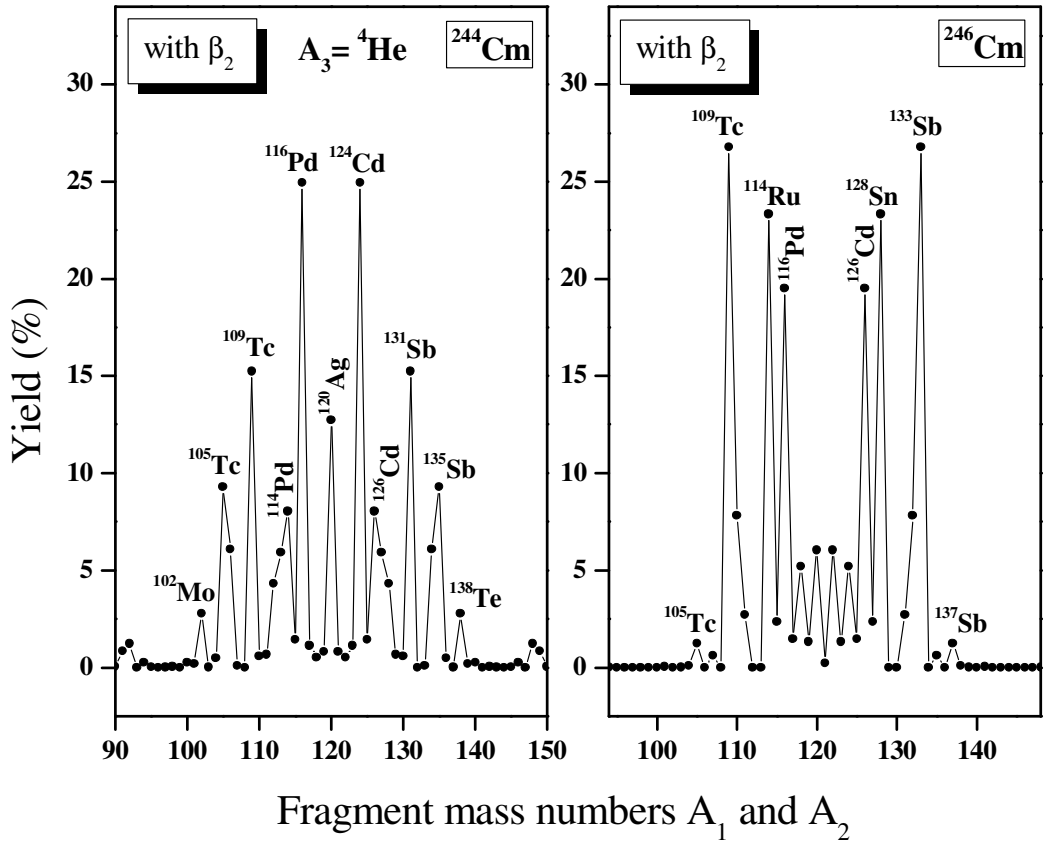

**Figure 15.** The calculated yields for the charge minimized third fragment $^4$He with the inclusion of quadrupole deformation $\beta_2$ plotted as a function of mass numbers $A_1$ and $A_2$ for $^{244-246}$Cm isotopes. The fragment combinations with higher yields are labeled.

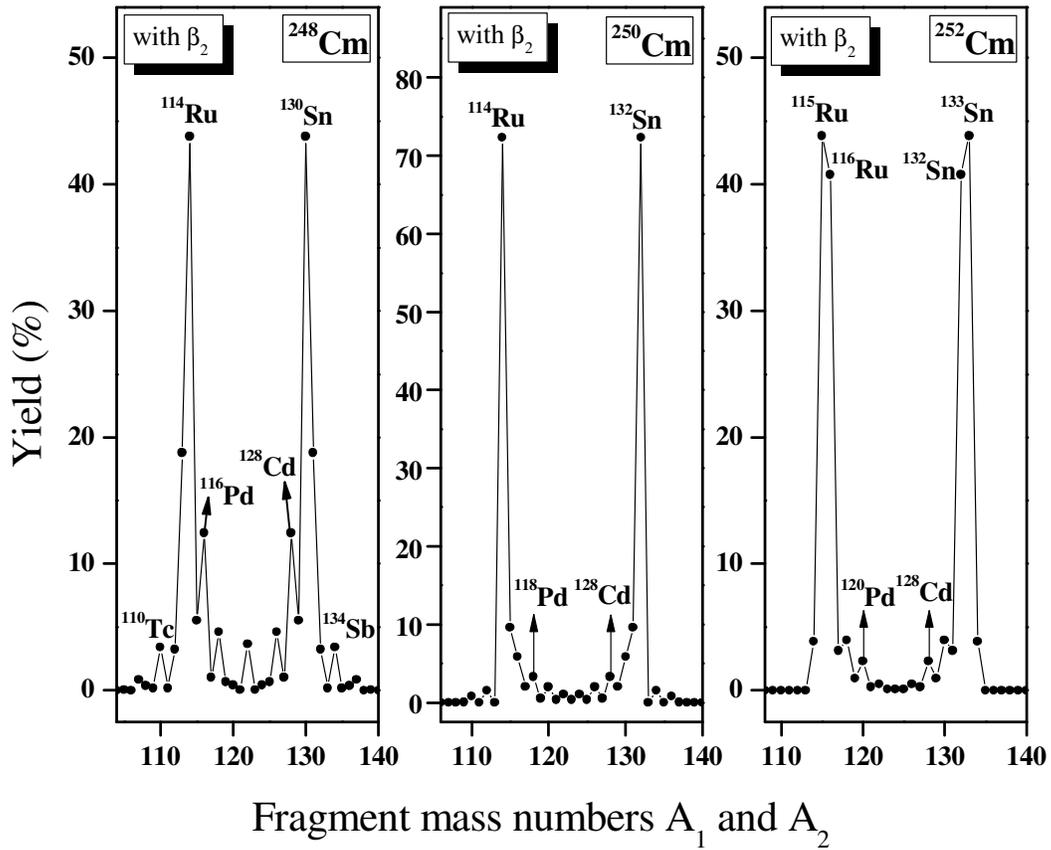

**Figure 16.** The calculated yields for the charge minimized third fragment $^4$He with the inclusion of quadrupole deformation $\beta_2$ plotted as a function of mass numbers $A_1$ and $A_2$ for $^{248-252}$Cm isotopes. The fragment combinations with higher yields are labeled.

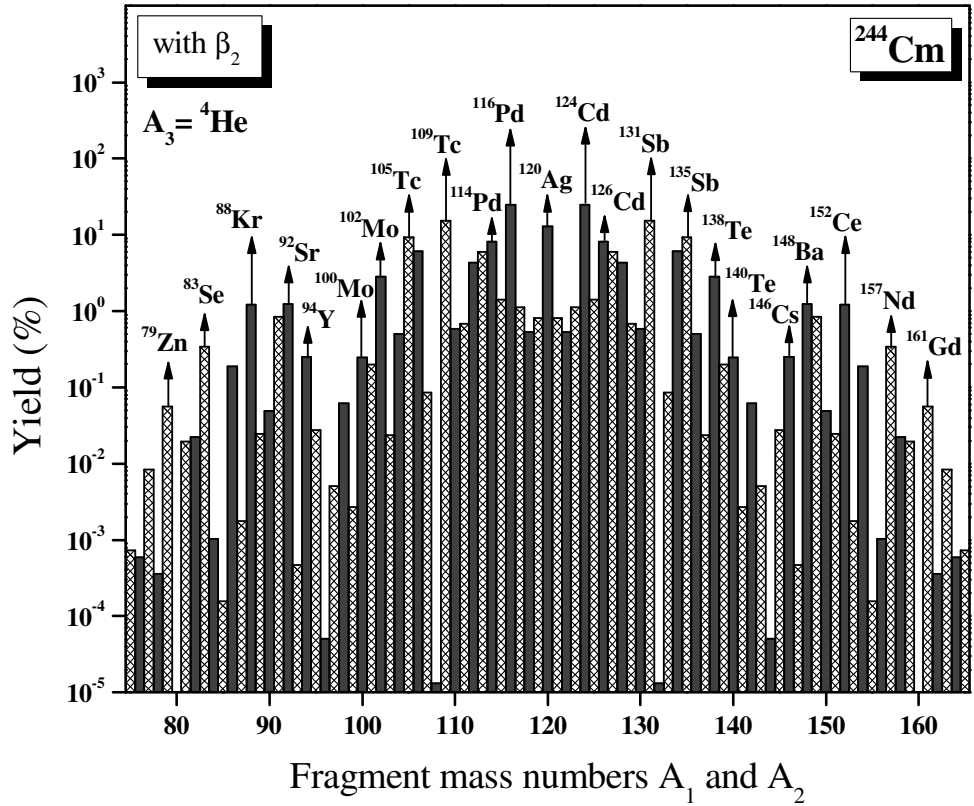

**Figure 17.** The calculated yields for the charge minimized third fragment $^4$He with the inclusion of quadrupole deformation $\beta_2$ plotted as a function of mass numbers $A_1$ and $A_2$ for $^{244}$Cm isotope.

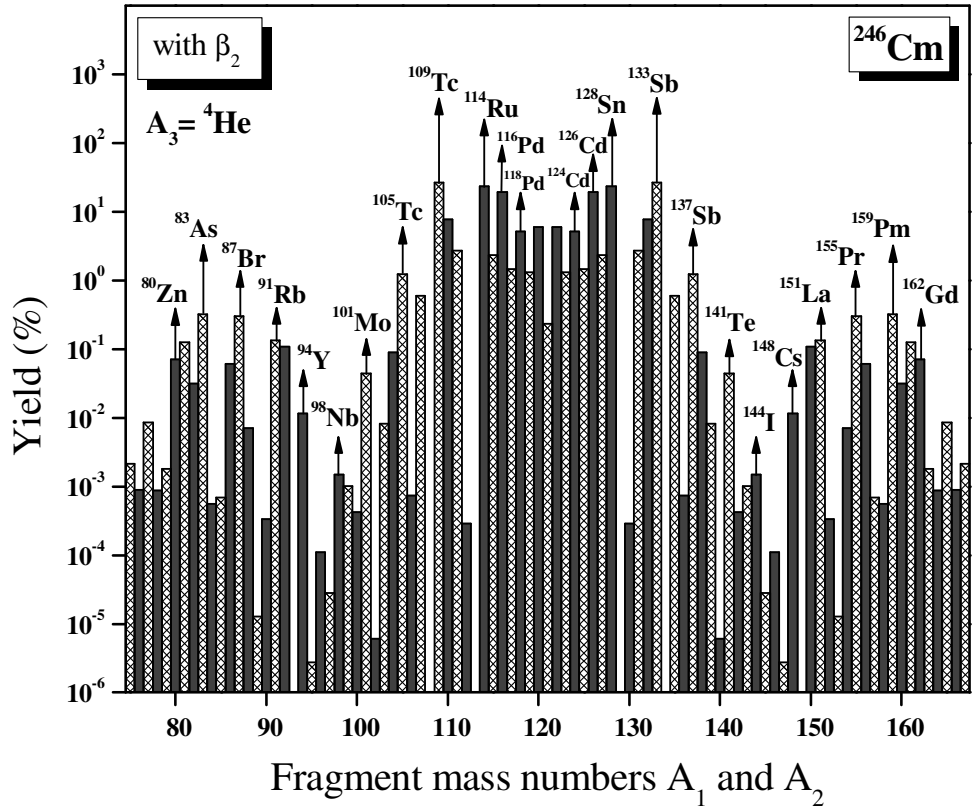

**Figure 18.** The calculated yields for the charge minimized third fragment $^4$He with the inclusion of quadrupole deformation $β_2$ is plotted as a function of mass numbers $A_1$ and $A_2$ for $^{246}$Cm isotope.

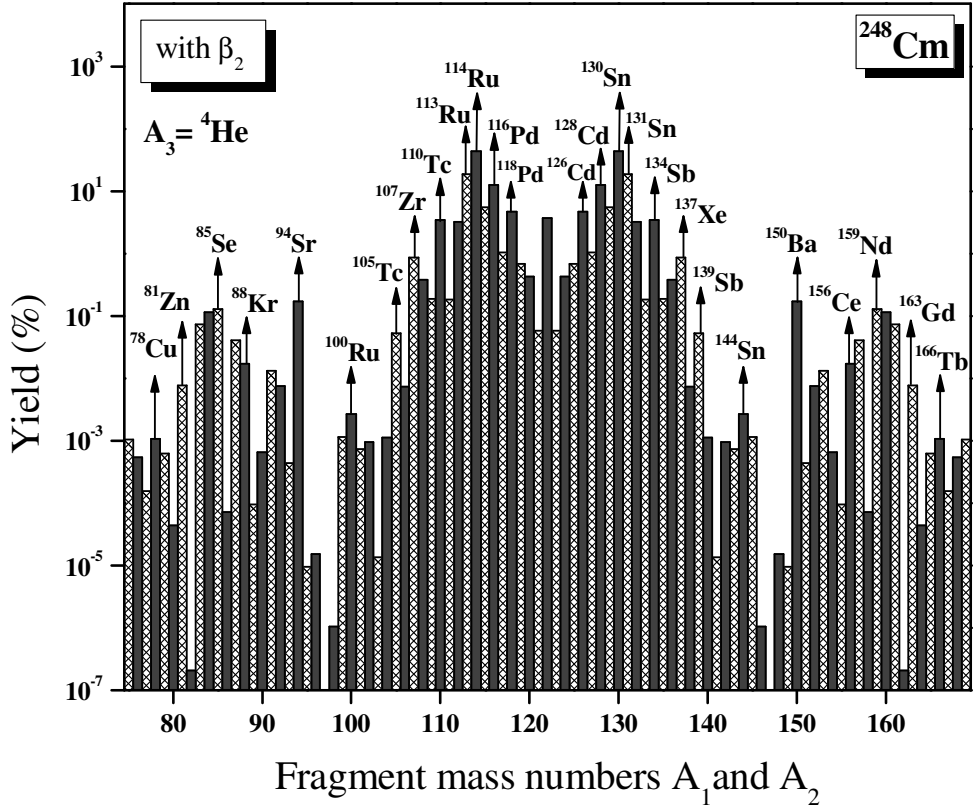

**Figure 19.** The calculated yields for the charge minimized third fragment ⁴He with the inclusion of quadrupole deformation $\beta_2$ plotted as a function of mass numbers $A_1$ and $A_2$ for $^{248}$Cm isotope.

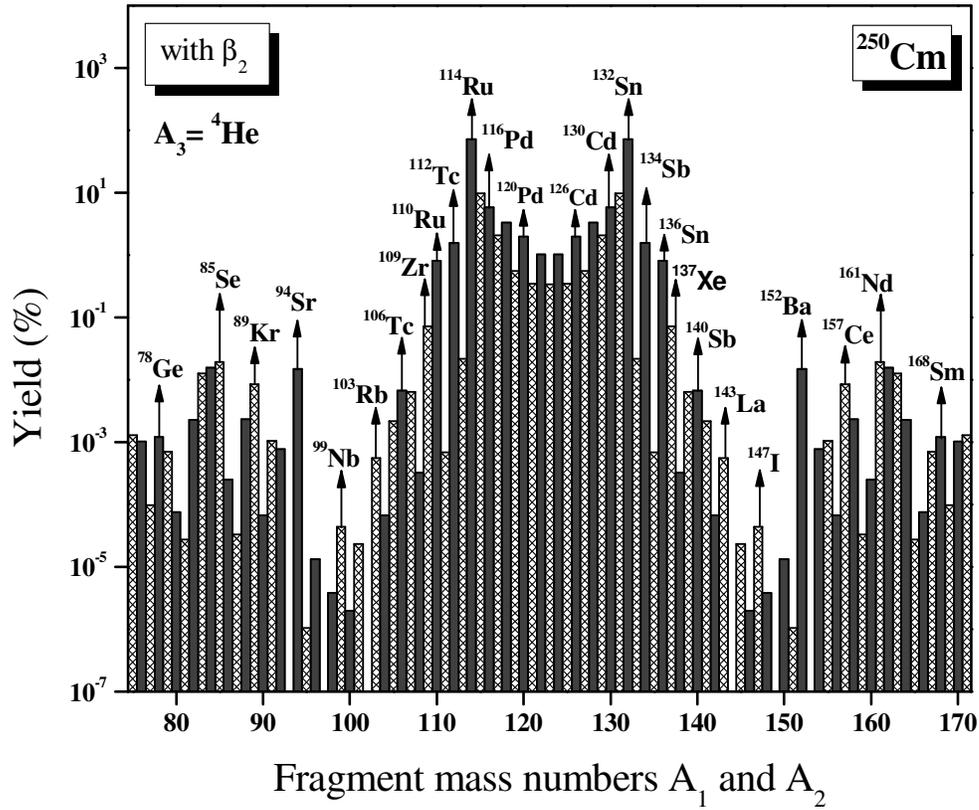

**Figure 20.** The calculated yields for the charge minimized third fragment $^4$He with the inclusion of quadrupole deformation $\beta_2$ plotted as a function of mass numbers $A_1$ and $A_2$ for $^{250}$Cm isotope.

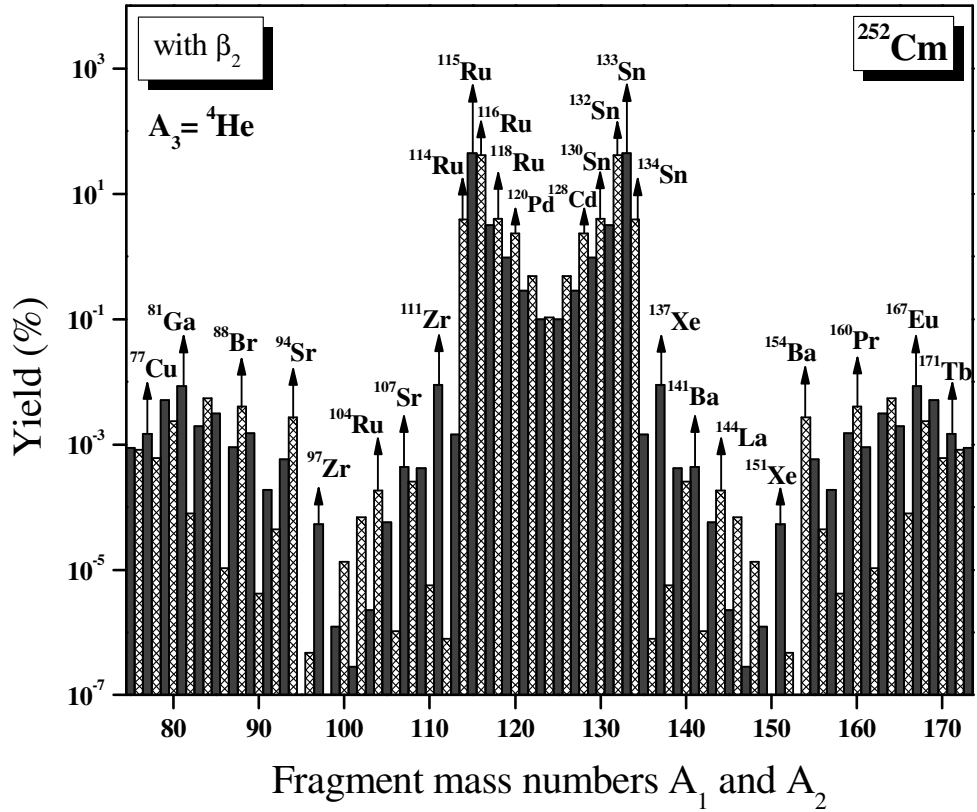

**Figure 21.** The calculated yields for the charge minimized third fragment $^4$He with the inclusion of quadrupole deformation $\beta_2$ plotted as a function of mass numbers $A_1$ and $A_2$ for $^{252}$Cm isotope.